\documentclass[aps,pre,reprint,superscriptaddress]{revtex4-2}
\usepackage[american]{babel}
\usepackage{bm}
\usepackage{amsfonts}
\usepackage{epsfig,graphicx}
\usepackage{amsmath}
\usepackage{amssymb}
\usepackage{amsthm}
\newcommand{\BE}{\begin{equation}}
\newcommand{\EE}{\end{equation}}
\newcommand{\corrLLL}{}

\newcommand{\mR}{\mathbb{R}}
\newcommand{\e}{\mathrm{e}}

\newcommand{\scalh}{\overline{h}}

\newcommand{\abs}[1]{{\vert {#1} \vert}}

\newcommand{\pt}{\partial}

\newcommand{\der}[2]{\dfrac{\pt #1}{\pt #2}}

\usepackage{color}
\usepackage{comment}

\newcommand{\bit}{\begin{itemize}}
\newcommand{\eit}{\end{itemize}}
\newcommand{\ben}{\begin{enumerate}}
\newcommand{\een}{\end{enumerate}}
\newcommand{\beq}{\begin{equation}}
\newcommand{\eeq}{\end{equation}}
\newcommand{\beqa}{\begin{eqnarray}}
\newcommand{\eeqa}{\end{eqnarray}}
\newcommand{\beqn}{\begin{eqnarray*}}
\newcommand{\eeqn}{\end{eqnarray*}}

\begin{document}

%\preprint{}

\title{Wigner function with correlation damping}


\author{Luigi Barletti}
\email{luigi.barletti@unifi.it}
%\thanks{}
\affiliation{Dipartimento di Matematica e Informatica ``U. Dini'', Universit\`a degli Studi di Firenze, Viale Morgagni 67/a, 50134 Firenze, Italy}
\author{Paolo Bordone}
\email{paolo.bordone@unimore.it}
%\thanks{}
\affiliation{Dipartimento di Scienze Fisiche, Informatiche e Matematiche, Universit\`a di Modena e Reggio Emilia, Via Campi 213/A, 41125 Modena, Italy}
\author{Lucio Demeio}
\email{l.demeio@univpm.it}
%\thanks{}
\affiliation{Dipartimento di Ingegneria Industriale e Scienze Matematiche, Universit\`a Politecnica delle Marche, Via Brecce Bianche 12, 60131 Ancona, Italy}
\author{Elisa Giovannini}
\email{elisa.giovannini@unifi.it}
%\thanks{}
\affiliation{Dipartimento di Matematica e Informatica ``U. Dini'', Universit\`a degli Studi di Firenze, Viale Morgagni 67/a, 50134 Firenze, Italy}


\date{\today}

\begin{abstract}
We examine  the effect of the decoherence-induced reduction of correlation length  on a one-dimensional scattering
problem by solving numerically the evolution equation
for the Wigner function with decoherence proposed by  Barletti {\it et al.} [Journal of Computational and Theoretical Transport {\bf 47}, 209 (2018)]. The numerical solution is achieved by the Splitting-Scheme algorithm, {\corrLLL suitably modified to include the decoherence term}. Three cases are examined, corresponding to a reflection-dominated regime, a transmission-dominated regime and an intermediate one. The dynamic evolution of the Wigner function is followed until the separation process of the reflected and of the transmitted packets is complete and it is observed for three different values of the correlation length. The outcomes show a broadening and flattening of the Wigner function which becomes progressively more pronounced as the correlation length is decreased. This results in a reduced reflection at low energies and in a reduced transmission at high energies.
\end{abstract}

%\keywords{}

\maketitle

%\section{}
\section{Introduction \label{S1}}
The Wigner-function (WF) approach to quantum kinetic theory was introduced in 1932 by Wigner \cite{Wigner32} 
(see also \cite{Tatarski83,Hillery84,Zachos05}) in order to calculate the quantum corrections
to thermodynamic equilibrium. It has been widely studied in the last four decades by many research groups, including mathematicians, physicists and engineers \cite{Wigner32,Kluksdahl89,Frensley90,ref1}. This approach relies on the phase-space representation of Quantum Mechanics, whose physical implications and mathematical properties have been analyzed in detail and are now better understood. There are, however, several practical and theoretical limitations to the Wigner formalism which make the applications difficult, so that only very few  results concerning real systems have been obtained so far by this method; in particular, a large-scale use in practical cases still looks to be a distant goal. These difficulties arise from some of the 
fundamental assumptions under which the Wigner equation holds, among which we recall: the vanishing of the density matrix $\rho(r,s)$ as $r, s \to \pm\infty$
(this allows the vanishing of the integrated terms in an integration by parts) for normalizable states; that the potential $V(x)$ must be defined over the whole
space; and no mechanism which destroys the phase correlations
of the individual states must be present, i.e., the correlation
length must be infinite.  These conditions entail that the WF formalism in its standard formulation is applicable only to fully Hamiltonian, spatially infinite coherent
systems, which are also confined. They also imply, for example, that boundary conditions cannot be included in the formalism without further theoretical modifications. Finally, the absence of decoherence mechanisms is one of the main causes of the highly oscillatory character of the Wigner function, which makes numerical simulations very difficult. Another issue related to the Hamiltonian, coherent character of the model is that the evolution equation for the Wigner function is invalid for scattering states, since the wave function does not vanish at infinity \cite{nedjalkov}. \\

In this paper we investigate the question related to the coherence length, by presenting  numerical simulations on a simple reflection-transmission problem approached by the decoherence model  introduced in \cite{Barletti18}. The results confirm the broadening and flattening of the Wigner function with time as predicted in \cite{Barletti18};  they also indicate that a reduced coherence length favors transmission of low-energy electrons through the potential barrier, inhibiting reflection. We observe another important feature, which appears when the potential varies abruptly, namely the formation of a narrow region of sharp variation of the Wigner function when the packet separation begins; our analysis strongly suggests an interpretation in terms of a more classical-like behavior as the correlation length is reduced.  \\

The Wigner function is the Wigner-Weyl transform of the density matrix $\rho(r,s)$, namely
\begin{equation}
\label{WWT}
  f(x,p, t) = \frac{1}{2\pi\hbar} \int \rho\left(x + \frac{\eta}{2},  x - \frac{\eta}{2},t \right) \e^{-i\eta p /  \hbar}\, d\eta 
\end{equation}
and it obeys the well-known governing evolution equation
\begin{equation}
\label{WE}
   \frac{\pt f}{\pt t} + \frac{p}{m} \frac{\pt f}{\pt x} + \frac{i}{\hbar}\varTheta[\delta V] f = 0,
\end{equation}
often called ``Wigner equation,'' where $\Theta[\delta V]$ is the pseudo-differential operator
\begin{eqnarray}
\label{Theta}
&& \left( \varTheta[\delta V]f\right)(x,p) = \nonumber \\
&& \quad = \frac{1}{2\pi\hbar}
    \int \int \delta V(x,\eta)\, f(x,p')\, \e^{i\eta (p'-p)/\hbar}\,dp' \,d\eta
\end{eqnarray}
where $V(x)$ is the potential energy
and  the symbol $\delta V$ is given by $\delta V(x,\eta) = V(x + \eta/2) - V(x - \eta/2)$.
In this work we focus on the modifications which must
be introduced in the WF formalism when the effects
of phase-randomization have to be considered. This is
an important factor, for example, in the description of
semiconductor devices with the WF approach \cite{Kluksdahl89,Frensley90,Querlioz10}.
This issue has been addressed in \cite{Ferrari06,Jacoboni,Schwaha13,Ellinghaus,Barletti18}. 
In \cite{Ferrari06} the finite size of a semiconductor device was considered by modifying
the correlation function across the device boundaries,
allowing in fact for a finite correlation length,
within the framework of the Schr\"odinger equation.
The results of that paper showed that the  finiteness of the
device or, equivalently, a finite correlation length favors
transmission of low-energy electrons through the potential
barrier in the scattering process, inhibiting reflection,
since long-wavelength components of the potential
cannot interfere effectively with the electron wave
function. 
In \cite{Jacoboni} a general exponential correlation damping
factor was introduced in the definition of the Wigner function,
which led to a modified evolution equation where, because
of the non-differentiable nature of the damping
factor, a complex momentum and a complex Wigner function were introduced.
In \cite{Barletti18} a Wigner equation with decoherence was introduced, the main effect of the decoherence being 
the decay in time of the correlation length. Such model can be considered  as the dynamical version of the approach of Ref. \cite{Jacoboni}.
\\

In this paper, we examine again the effect of the finite 
correlation length on a one-dimensional scattering
problem by solving numerically the evolution equation
for the Wigner function proposed in \cite{Barletti18}. 
A Gaussian wave packet, supposed free at $t \to -\infty$, enters a region
where a Gaussian shaped potential is present, and
the final state is observed in terms of average quantities
(momentum and position), transmission coefficient 
and density profiles.  The numerical solution is obtained
with the splitting-scheme algorithm \cite{Cheng76,suh,Arnold95,Arnold96,Demeio03}, suitably
modified in order to accommodate for the extra terms
which arise because of the finite correlation length.
The paper is organized as follows. In Sec. \ref{sec:WignerDecoherence} we introduce
the Wigner equation with decoherence and describe its properties;
Sec. \ref{sec:PhysicalModel} briefly introduces the physical model; in Sec. \ref{sec:NumericalMethod} we illustrate the numerical method;
Sec. \ref{sec:NumericalResults} contains the numerical results and in Sec.
\ref{sec:Conclusions} we state our conclusions.

\section{Wigner equation with decoherence} \label{sec:WignerDecoherence}
In order to endow the WF formalism with a  
mechanism describing decoherence, a model was developed by Barletti {\it et al.} in  \cite{Barletti18}, based on the rigorous results of Adami {\it et al.} \cite{Adami03}. The idea is to let the carriers, described by the WF formalism, undergo a number of collisions per unit time with a nominal background medium of light particles; each interaction is described by the model introduced in \cite{Adami03} and, in the limit of very small mass ratio, it amounts to the following transformation of the particle density matrix
\begin{equation}
\label{interaction}
  \rho(x,y,t_0) \longmapsto \mathcal{I}(x,y) \rho(x,y,t_0).
\end{equation}
Here, each collision is supposed to be instantaneous, $t_0$ is the time at which the collision occurs and
\begin{equation}
\label{Idef}
  \mathcal{I}(x,y) = \varDelta_\lambda(x-y) + i\varGamma(x) - i\varGamma(y),
\end{equation}
where $\varDelta_\lambda$ and $\varGamma$ are quantities which depend on the light particle wave function and on the scattering coefficients
\cite{Barletti18,Adami03}.
In particular, $\varDelta_\lambda(\eta)$ describes the damping of the correlation for large values of $x-y$; it depends on the positive parameter $\lambda$ which is the typical length of the correlation damping (we will often refer to it as ``correlation length''), with $\lambda \to +\infty$ for the fully coherent system. We also assume that
 \beqa
 && \varDelta_\lambda(0) = 1 \label{Delta0} \\
&& \lim_{\eta \to \pm \infty} \varDelta_\lambda(\eta) = 0 \label{Deltainf} \\
&& \lim_{\lambda \to  \infty} \varDelta_\lambda(\eta) = 1. \label{Deltacoh}
\eeqa
\noindent In \cite{Jacoboni}, the function $\varDelta_\lambda(x-y) = \e^{-\abs{x-y}/\lambda}$, which fulfills all these requirements, was chosen. \\

In the WF formalism Eq. \eqref{interaction} becomes \cite{Barletti18}
\begin{equation}
  f \longmapsto \widehat\varDelta_\lambda \ast f + \frac{i}{\hbar}\varTheta[\delta \varGamma] f,  \label{Wignercorr}
\end{equation}
where $\ast$ denotes convolution with respect to the momentum $p$ and 
$$
 (\widehat \varDelta_\lambda)(p) = \frac{1}{2\pi\hbar} \int_\mR \varDelta_\lambda(\eta) \e^{-i\eta p /  \hbar}\, d\eta
$$
is the Fourier transform of $\varDelta_\lambda$.
From Eq. (\ref{Wignercorr}) we see that $\varGamma$ plays the role of a potential term and, therefore, it does not contribute to decoherence. 
In addition, $\varGamma$ is usually small \cite{Adami03} and we shall neglect its contribution.
Assuming that collisions occur randomly with frequency $1/\tau_0$, we obtain the following equation for the Wigner function (see \cite{Barletti18} for the details) 
\begin{equation}
\label{WED}
   \frac{\pt f}{\pt t} + \frac{p}{m} \frac{\pt f}{\pt x} + \frac{i}{\hbar}\varTheta[\delta V] f = \frac{1}{\tau_0}\,( f_\lambda - f),
\end{equation}
where
\begin{equation}
f_\lambda(x,p,t) = (\widehat\varDelta_\lambda \ast f)(x,p,t) =  \int_\mR \hat\varDelta_\lambda(p-p') \,f(x,p',t) \, dp'. \label{fD}
\end{equation}

The collisional term at the right-hand side of Eq. \eqref{WED} comes from the interactions with the environment and represents therefore a decoherence mechanism.
Note that this term describes the relaxation of $f$ to a modified Wigner function $f_\lambda$,
which is exactly the Wigner function with finite coherence length defined by Jacoboni and Bordone \cite{Jacoboni}.
To this extent, we can interpret \eqref{WED} as the dynamical version of the model introduced in \cite{Jacoboni}. As shown in  \cite{Barletti18}, when $\Delta_\lambda(\eta)$ is regular enough it admits the following expansion:
\begin{equation}
\label{expansion}
   \Delta_\lambda(\eta) = 1 + i\Lambda_1\eta - \Lambda_2\eta^2 + \cdots,
\end{equation}
where $\Lambda_1$ and $\Lambda_2$ are real, with $\Lambda_2 > 0$. 
We remark that, if $\Lambda_1 = 0$, the quadratic approximation 
$\varDelta_\lambda(\eta) \approx 1 - \varLambda_2 \,\eta^2$ reduces \eqref{WED} to the 
Wigner-Fokker-Planck equation
\begin{equation}
\label{WFP}
   \frac{\pt f}{\pt t} + \frac{p}{m} \frac{\pt f}{\pt x} + \frac{i}{\hbar}\varTheta[V] f =   \frac{\hbar^2\varLambda_2}{\tau_0} \frac{\pt^2 f}{\pt p^2},
\end{equation}
which is a classic model of decoherence \cite{Joos85}.
\\

As with the standard WF formalism, the macroscpic quantities are given by the moments of the Wigner function. In particular, for a given Wigner function $f$, the associated macroscopic density $n[f]$, current $j[f]$ and energy density $e[f]$ 
correspond to the first three moments of $f$, namely:
\beqa
  && n[f](x,t) = \int f(x,p,t)\,dp, \label{densf}
\\
  && j[f](x,t) = \frac{1}{m} \int p\,f(x,p,t)\,dp, \label{currf}
\\
 && e[f](x,t) = \frac{1}{2m} \int  p^2 \,f(x,p,t)\,dp.\label{enerf}
\eeqa
When the same quantities are evaluated with $f_\lambda$, by
using \eqref{expansion} it can be easily proven that
\beqa
  && n[f_\lambda] = n[f], \label{densflam}
\\
   && j[f_\lambda] = j[f] + \frac{\hbar\Lambda_1}{m}\, n[f],  \label{currflam}
\\
   && e[f_\lambda] = e[f] + \frac{\hbar\Lambda_1}{m}\, j[f] + \frac{\hbar^2\Lambda_2}{m}\, n[f].  \label{enerflam}
\eeqa
Hence the decoherence mechanism, as represented by the collision term in Eq. (\ref{WED}), conserves the density, while it modifies the current if and only if $\Lambda_1 \not= 0$ (which corresponds to the 
fact that, in this case, the background medium has an asymmetric distribution in $p$, so that an average nonzero momentum is transferred to the particle).
Note also that, since $\Lambda_2 > 0$, energy is absorbed from the environment at a rate $\hbar^2\Lambda_2/(m\tau_0) \, n$. \\

One word of caution should be issued before we present the physical model and our numerical results. The decoherence mechanism introduced in \cite{Barletti18} results in a progressive broadening and flattening of the Wigner function with time, this effect being stronger at small correlation lengths. This leads to an unphysical behavior for large times, already pointed out in \cite{Joos85}. However, in this paper we apply the finite correlation model to a scattering process, following the time evolution only up to the separation of the packets, and the long-time behavior of the system will not be followed. In our simulation, this unphysical behavior is revealed by the results for the smaller value of $\lambda$ considered, and will be commented upon in Sec. \ref{sec:NumericalResults}.

\section{The physical model} \label{sec:PhysicalModel}
In Sec. \ref{sec:NumericalMethod}, we shall present the numerical solution of Eq. \eqref{WED} for the case of a minimum uncertainty Gaussian wave packet which approaches 
a central region where the Gaussian potential
\begin{equation}
  V(x) = V_0 \, \e^{-x^2/a^2} \label{potential}
\end{equation}
is present. The model is similar to the one used in {\corrLLL \cite{Turner87}}. \\

The initial condition for Eq. \eqref{WED} is given by the Wigner-Weyl transform of an initial wave function
of the form
$$
  \psi(x,0) = \sqrt[4]{\frac{2\sigma_p}{\pi\hbar^2}} \exp \left\{ - \frac{\sigma_p^2(x-x_0)^2}{\hbar^2} + i\,\frac{p_0(x-x_0)}{\hbar}\right\},
$$
where the normalization constant is such that $\int \abs{\psi(x)}^2 dx = 1$.
Here, $\sigma_p$ is the initial momentum spread, $x_0$ the initial average position and $p_0$ the initial average momentum.
The initial Wigner function can be easily calculated from Eq. \eqref{WWT} and is given by
\begin{equation}
\label{IVF}
  f(x,p,0) = \frac{1}{\pi\hbar} \exp \left\{ - 2\, \frac{\sigma_p(x-x_0)^2+\sigma_x(p-p_0)^2}{\hbar^2} \right\},
\end{equation}
where $\sigma_x$ is the initial position variance which, for a minimal uncertainty packet such as our $\psi(x,0)$, 
obeys the relationship $\sigma_p\sigma_x = \hbar^2/2$. The normalization of the wave function entails that the Wigner function is also normalized,
\beq
\int\int f(x,p,0) \, dx \, dp = 1.  \label{WFnorm}
\eeq
Since the Wigner equation (\ref{WED}) preserves the norm of the Wigner function, the normalization condition (\ref{WFnorm}) remains valid at all times. The initial Wigner function starts in free motion at a large distance from the origin; the exact initial condition is 
characterized by a dimensionless energy given by
\begin{equation}
\label{EK}
  E_K = \frac{p_0^2}{2mV_0}
\end{equation}
and an initial dimensionless momentum variance $\sigma_0 = \sigma_p/(mV_0)$. 
We also introduce the dimensionless variables $x' = x/a$, $t' =\sqrt{V_0/(ma^2)}$ and
the dimensionless Planck constant $\scalh = \hbar/(a\sqrt{mV_0})$, coherence length $\lambda' = \lambda/a$ 
and collision time $\tau = \tau_0/t'$; 
in the sequel, we shall drop the primes for a simpler notation and unprimed variables are meant to be dimensionless
from now on. 
The initial dimensionless momentum then is $p_0 = \sqrt{2E_K}$, while $x_0$ is chosen empirically so that the bulk 
of the initial packet lies far away from the influence region of the potential. 
The initial correlation variance is initially set to zero, but it acquires non-zero values during the time evolution.
With the dimensionless variables, Eq. \eqref{WED} becomes
\begin{equation}
\label{WED2}
   \frac{\pt f}{\pt t} + p\frac{\pt f}{\pt x} + \frac{i}{\scalh}\varTheta[\delta V] f = \frac{1}{\tau}\,( f_\lambda - f),
\end{equation}
where
\begin{equation}
f_\lambda(x,p,t) = \frac{1}{2\pi\scalh } \int_\mR \int \varDelta_\lambda(\eta) \e^{-i\eta (p-p') /  \scalh} \,f(x,p',t)\,d\eta \, dp',
\end{equation}
and the initial condition becomes
\begin{equation}
 f(x,p,0) = \frac{1}{\pi\scalh}e^{-2\sigma_0^2\left[(x-x_0)^2/\scalh^2+\scalh^2(p-p_0)^2\right]}.
\end{equation}
In the numerical simulations we shall mainly follow the average quantities and the particle density. With the Wigner function normalized according to (\ref{WFnorm}) we have
\beqa
&& <x>(t) = \int\int x \, f(x,p,t) \, dx \, dp \equiv x_t \label{average1} \\
&& \qquad \mbox{(average position)}  \nonumber \\
&& <p>(t) = \int\int p \, f(x,p,t) \, dx \, dp \equiv p_t \label{average2} \\
&& \qquad \mbox{(average momentum)}  \nonumber \\
&& \sigma_{20}(t) = \int\int (x-x_t)^2 \, f(x,p,t) \, dx \, dp \label{average3} \\
&& \qquad \mbox{(position spread or variance)}  \nonumber \\
&& \sigma_{02}(t) = \int\int (p-p_t)^2 \, f(x,p,t) \, dx \, dp \label{average4} \\
&& \qquad \mbox{(momentum spread or variance)}  \nonumber \\
&& \sigma_{11}(t) = \int\int (x-x_t) \, (p-p_t) \, f(x,p,t) \, dx \, dp.  \label{average5}  \\
&& \qquad \mbox{(covariance)} \nonumber
\eeqa
with the initial conditions $<x>(0)=x_0$, $<p>(0)=p_0$, $\sigma_{02}(0)=\sigma_0$, $\sigma_{20}(0)=\scalh^2/(2 \, \sigma_0)$,  $\sigma_{11}(0)=0$.

\section{The numerical method} \label{sec:NumericalMethod}

{\corrLLL We solve Eq. \eqref{WED2} by using {\corrLLL a modified version of }the splitting scheme algorithm, a 
 method which was initially developed in \cite{Cheng76} for the classical Vlasov equation for collisionless plasmas, and subsequently adapted to the quantum case  \cite{suh,Arnold95,Arnold96,Demeio03} in order to solve the Wigner equation \eqref{WE}.
\par
In its original formulation \cite{Cheng76} for the classical nonlinear Vlasov-Poisson system, the splitting scheme performs the numerical integration 
along the characteristics in the phase space.
A discretized mesh is introduced in the simulation domain of the phase space,
and the solution is advanced in time from $t$ to $t + \Delta t$
by alternating an integration along $x$ for half time-step (corresponding to integrating the equation with the free-streaming term only),
an integration along $p$ for a whole time step (corresponding to integrating the equation with the force term only) and a final integration again along $x$ for half time-step. The
integration along $x$ corresponds to a shift of the solution along $x$ (referred to as ``horizontal shift''), while
the integration along $p$ corresponds to a shift of the solution along $p$ (referred to  as ``vertical shift'').  At each time step after the initial one, the first horizontal shift can be
combined with the second horizontal shift of the previous time step in a unique horizontal shift, thus saving computational time.
\par

The quantum version of the splitting-scheme algorithm follows the same pattern \cite{suh,Arnold95,Arnold96,Demeio03}, even though there are no characteristics as in the classical case. 
The free-transport part of the equation is exactly the same as in the Vlasov case, while 
the force term of the Vlasov equation is replaced by the pseudo-differential operator term 
$i \, \varTheta[\delta V] f  / \scalh$. In our modified version of the splitting-scheme, we have added the decoherence term $- ( f_\lambda - f) / \tau$ to the pseudo-differential operator in the vertical shift.
In practical terms, in the vertical shift we solve the equation
\begin{equation}
\der{f}{t} + \frac{i}{\scalh}\varTheta[\delta V] f - \frac{1}{\tau}\,( f_\lambda - f)  =0
\end{equation}
which in the Fourier transformed space, becomes
\begin{equation}
\der{g}{t} + \frac{i}{\scalh} \, \delta V g - \frac{1}{\tau}(\varDelta_\lambda-1) g =0,
\end{equation} 
where
\begin{equation}
g(x,\eta,t) = \int f(x,p,t)e^{i\eta p/\scalh}dp.
\end{equation}
Hence,
\begin{equation}
g(x,\eta,t+{\Delta t})=e^{-\Delta t [i \, \delta V / \scalh +(1-\varDelta_\lambda)/\tau]} g \left(x,\eta,0\right).
\end{equation}
The phase-space integrals needed for the calculations of the average quantities (\ref{average1})-(\ref{average5}) are performed by standard open Newton-Cotes rules.}

\section{Numerical results} \label{sec:NumericalResults}
In this section we present and discuss the numerical solution of the Wigner equation with decoherence (\ref{WED})  for the physical model introduced in Sec. \ref{sec:PhysicalModel}. 
For the correlation-damping function $\varDelta_\lambda$ (introduced in Eq. (\ref{Idef})) we use 
\begin{equation}
\label{eqn:lambda_cosh}
\varDelta_\lambda(\eta) = \frac{1}{\cosh(\eta/\lambda)},
\end{equation}
which satisfies all properties outlined in Sec. \ref{sec:WignerDecoherence} and is differentiable over the whole domain. \\

\noindent We present the numerical results for four different values of $\lambda$, {\it (i)} $\lambda \to \infty$, corresponding to the quantum standard dynamics; {\it (ii)} $\lambda=10$, corresponding to a long correlation length;  {\it (iii)} $\lambda=4$, corresponding to an intermediate correlation length; and {\it (iv)} $\lambda=1$ corresponding to a short correlation length and for three different values of the dimensionless energy $E_K$, {\it (i)}  $E_K=0.5$ which corresponds to a reflection-dominated regime in the quantum standard dynamics; {\it (ii)} $E_K=1$, which corresponds to an intermediate reflection-transmission regime; and {\it (iii)} $E_K=1.5$, which corresponds to a transmission-dominated regime. Furthermore, we set $\tau=3$ in all simulations. For each value of the energy and of the correlation length we present the most relevant mean quantities as functions of time, i.e. the average position $<x>(t)$, the average momentum $<p>(t)$ and the position and momentum spreads $\sigma_{20}(t)$ and $\sigma_{02}(t)$, all defined in equations (\ref{average1})-(\ref{average4}). We also show the particle density $n(x)$ at four key instants of time and the Wigner function at the final time of the numerical simulation. Finally, we show the transmission coefficient $T$ as a function of the energy in the range $0.5 \le E_K \le 2$ for $\lambda \to \infty$, $\lambda=10$ and $\lambda=4$. For the transmission coefficient we adopt the approximate expression {\corrLLL \cite{Turner87}}
\begin{equation}
\label{eqn:T}
\displaystyle T=\frac{1}{2}\left( 1+\frac{\langle p\rangle_\infty}{p_0}\right),
\end{equation}
where $p_0$ is the initial momentum and $\langle p\rangle_\infty$ is the average momentum at the end of the simulation. 

\subsection{First case, reflection-dominated regime: $E_K=0.5$}

\begin{figure}[ht!]
\begin{center}
\epsfig{file=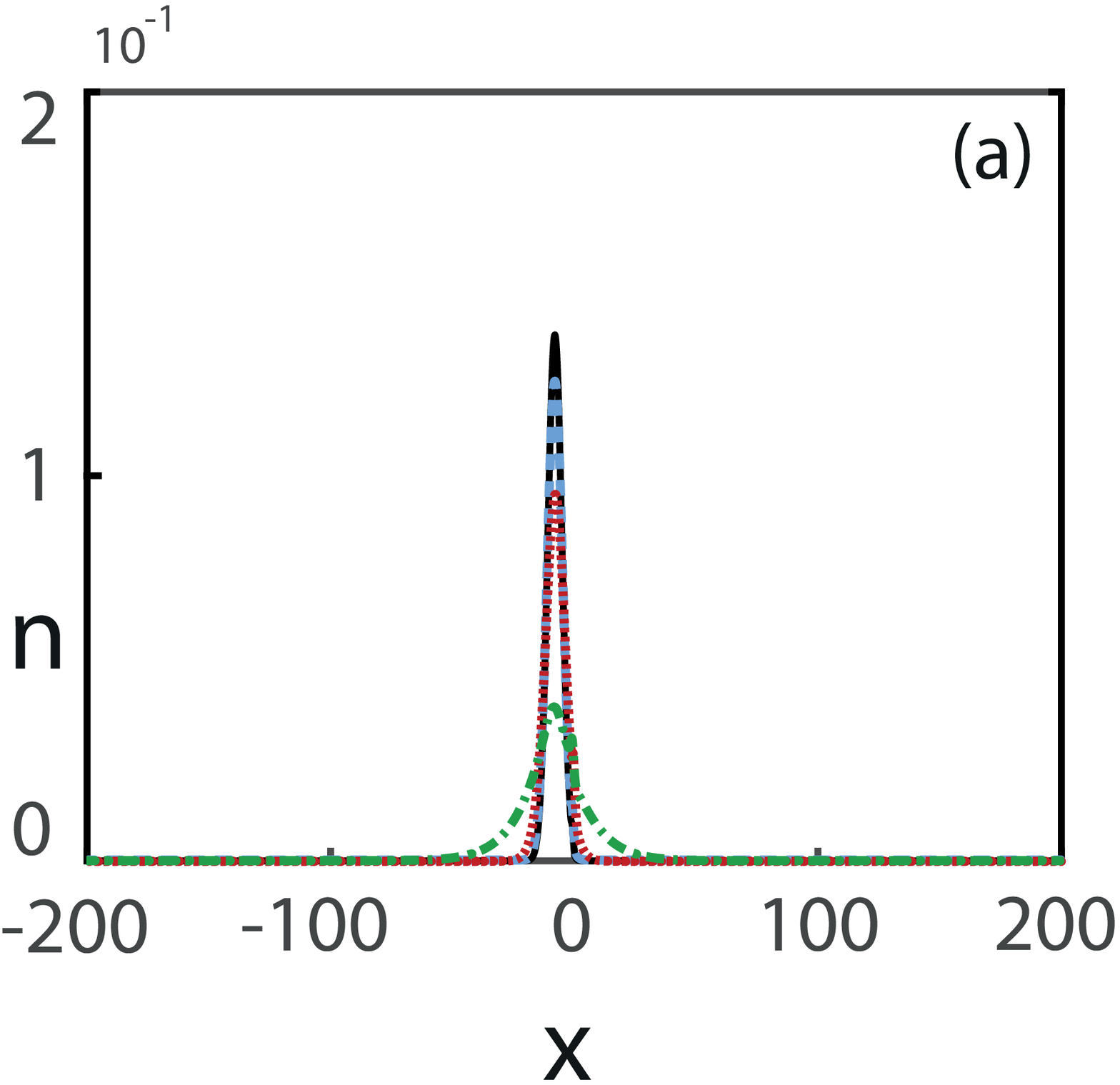,scale=0.24}\hspace*{0.1 cm}\epsfig{file=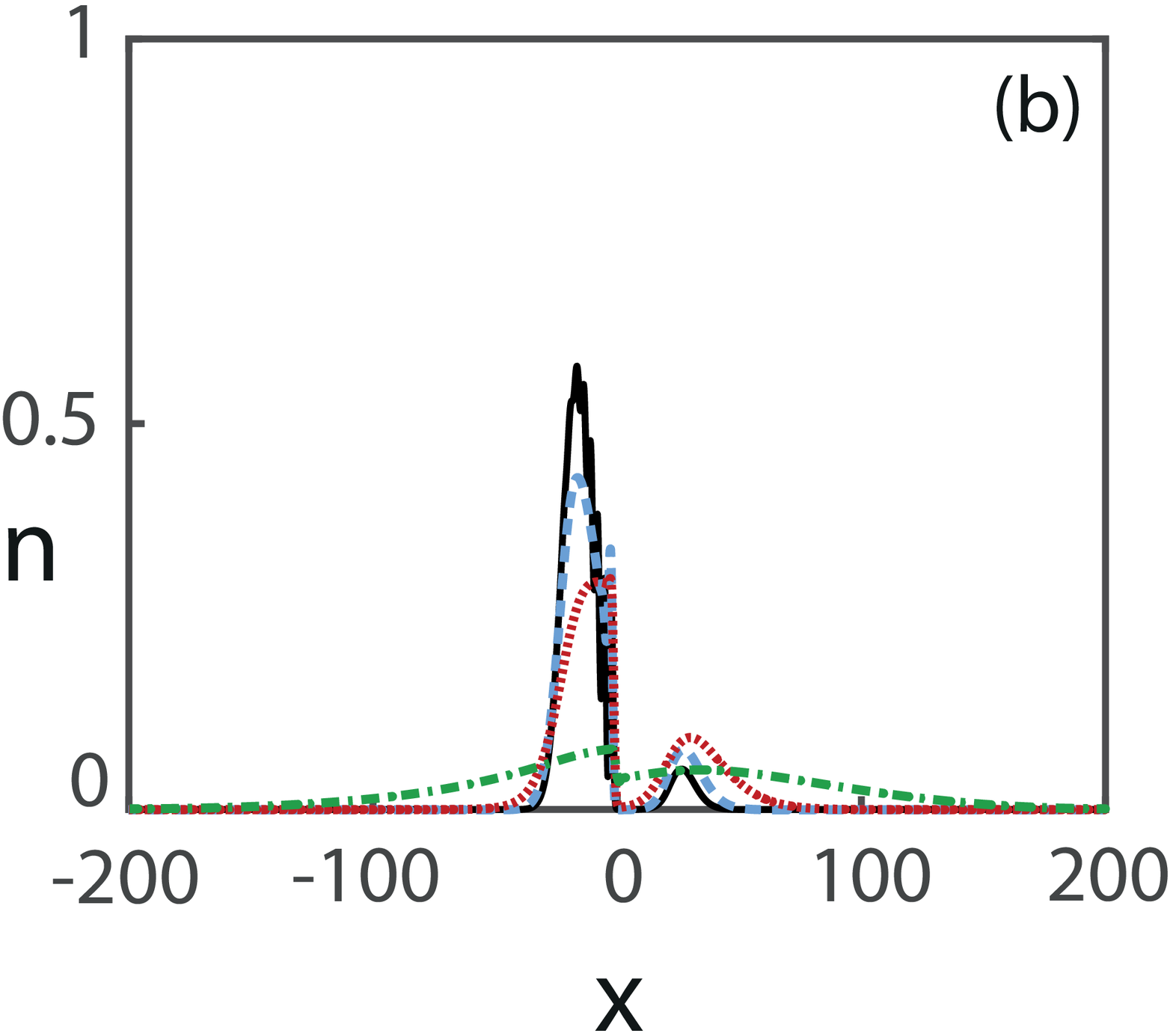,scale=0.24} 
\epsfig{file=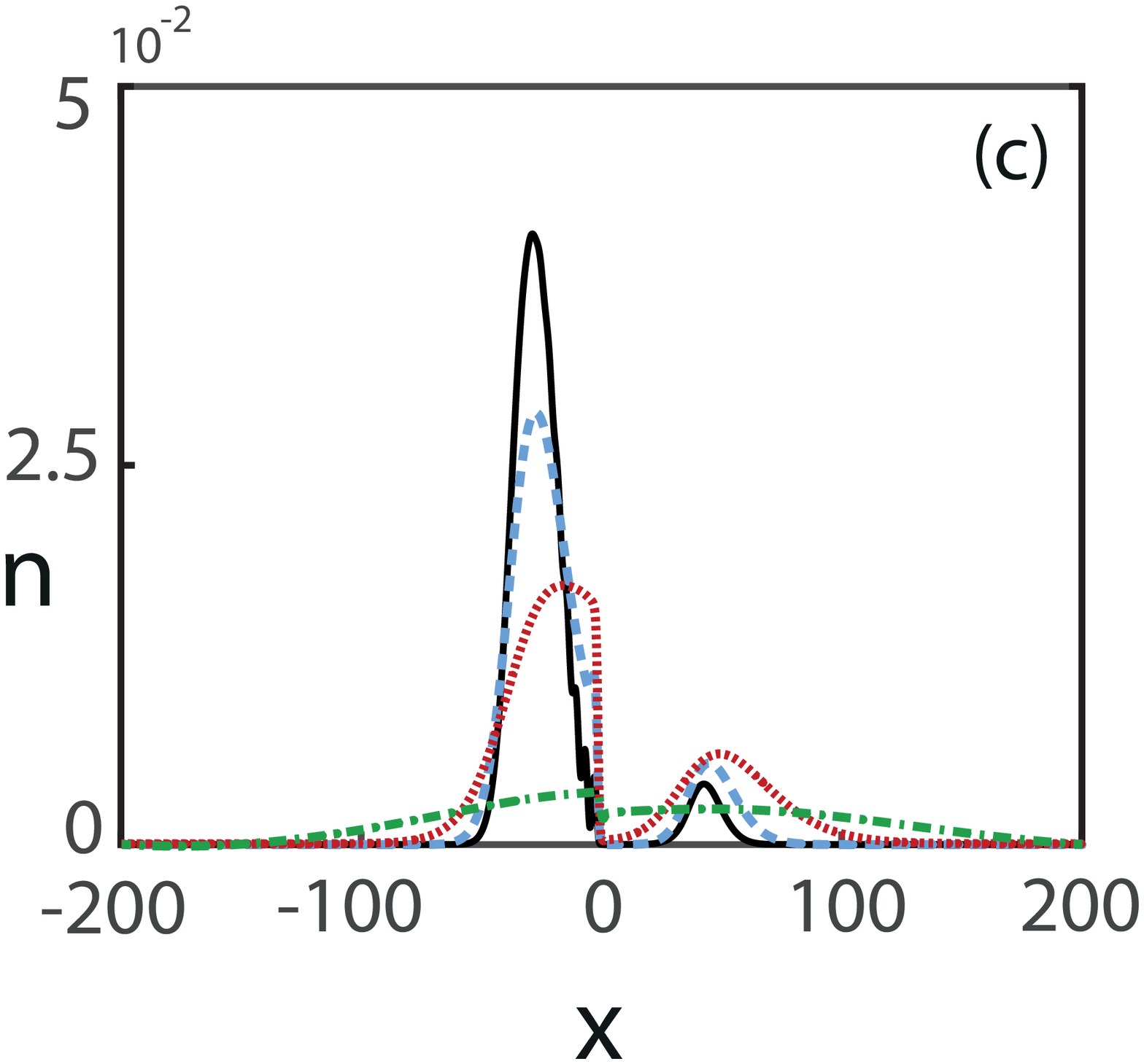,scale=0.24}\hspace*{0.1 cm}\epsfig{file=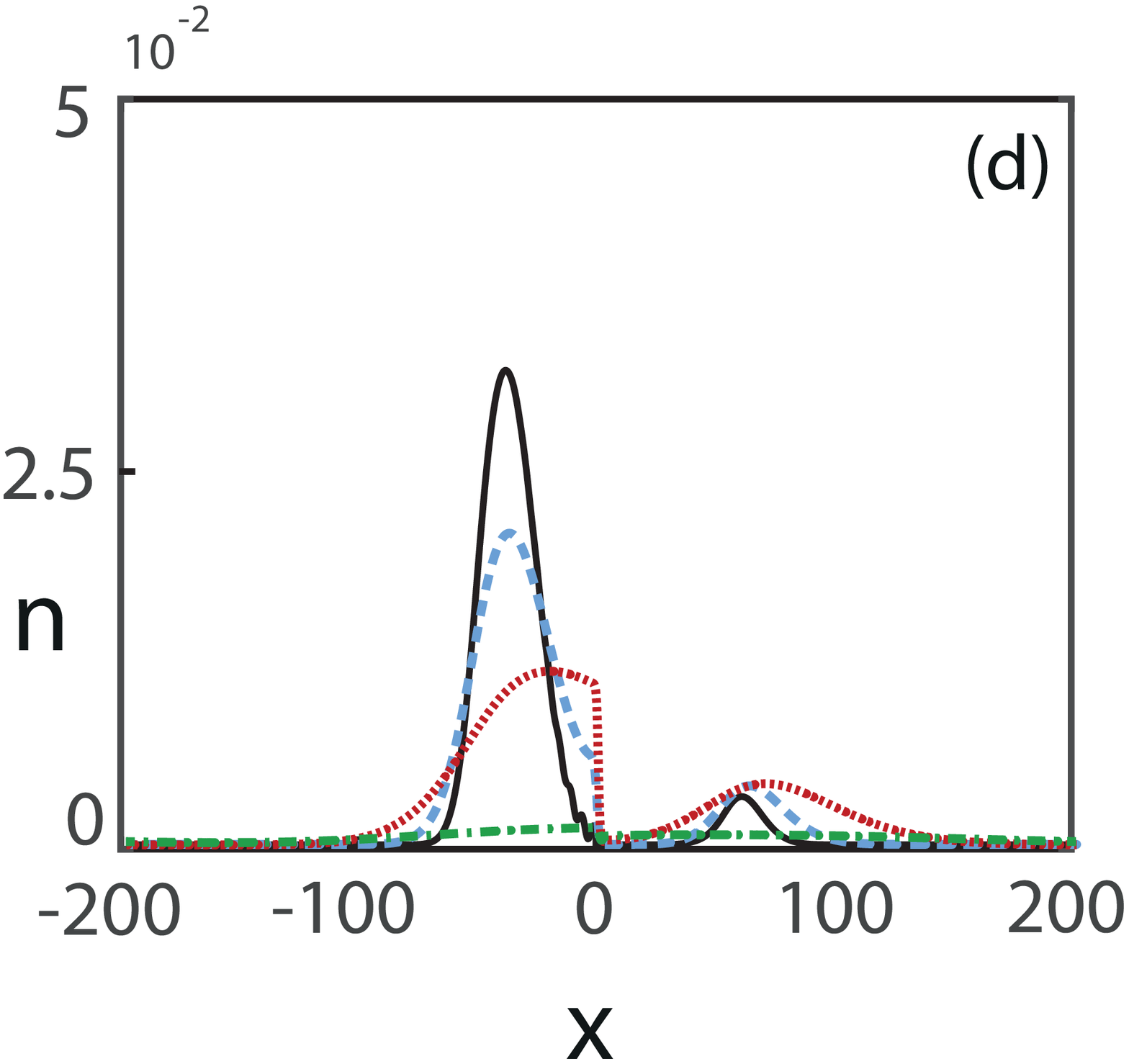,scale=0.24} 
\caption{\label{fig:dens05} Density profile $n(x)$ in dimensionless units as function of $x$ at four instants of time, $t=12$ (a), $t=36$ (b), $t=48$ (c), $t=60$ (d) for $E_K=0.5$, $\tau=3$, $\sigma_0=0.1E_K$ and four different values of $\lambda$. Coherent case ($\lambda\rightarrow \infty$): solid black line; $\lambda=10$: dashed blue line; $\lambda=4$: dotted red line;  $\lambda=1$: dash-dotted green line.} 
\end{center}
\end{figure}

\begin{figure}[ht!]
\setlength{\unitlength}{1cm}
\begin{center}
\epsfig{file=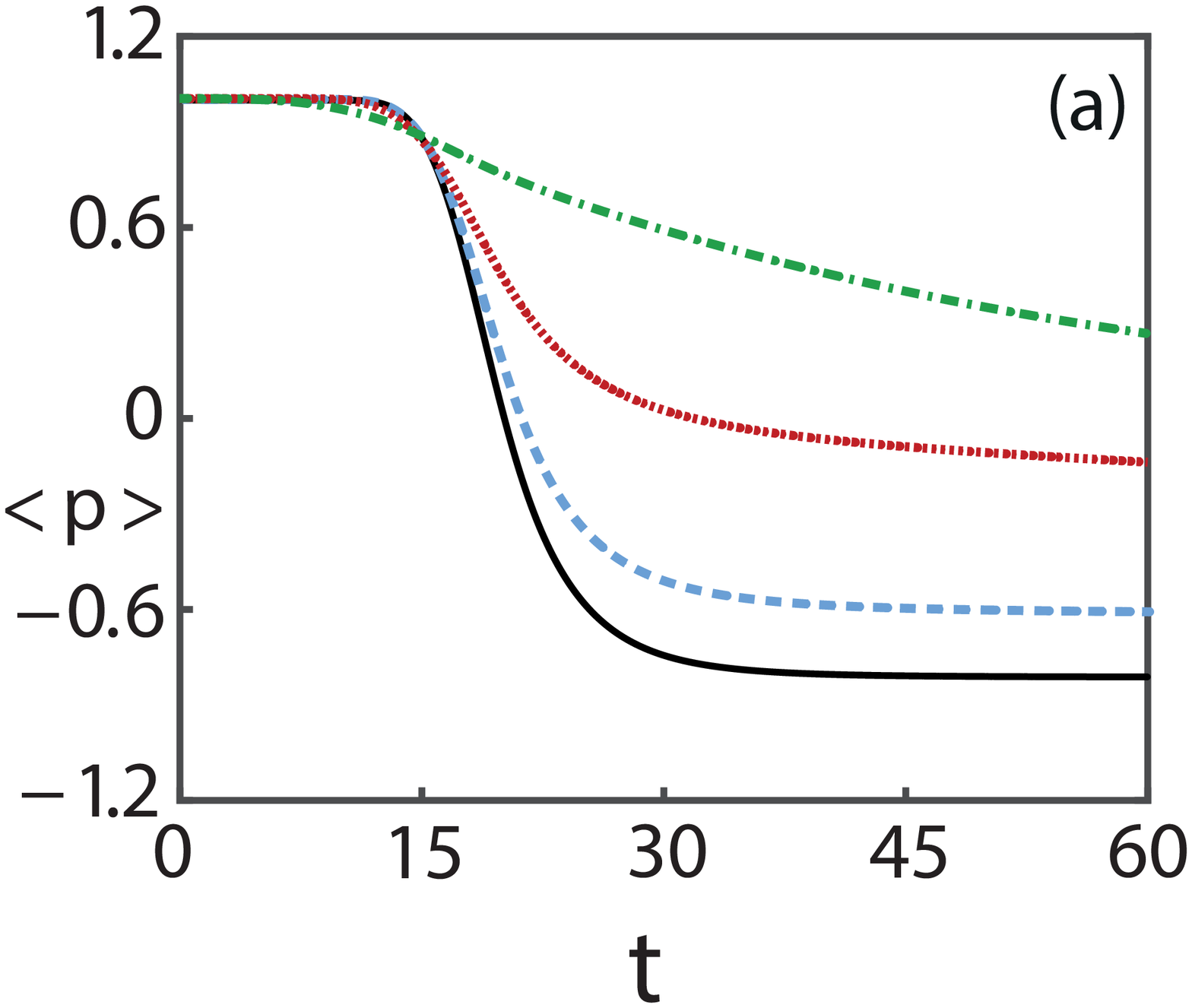,scale=0.24}\hspace*{0.1 cm}\epsfig{file=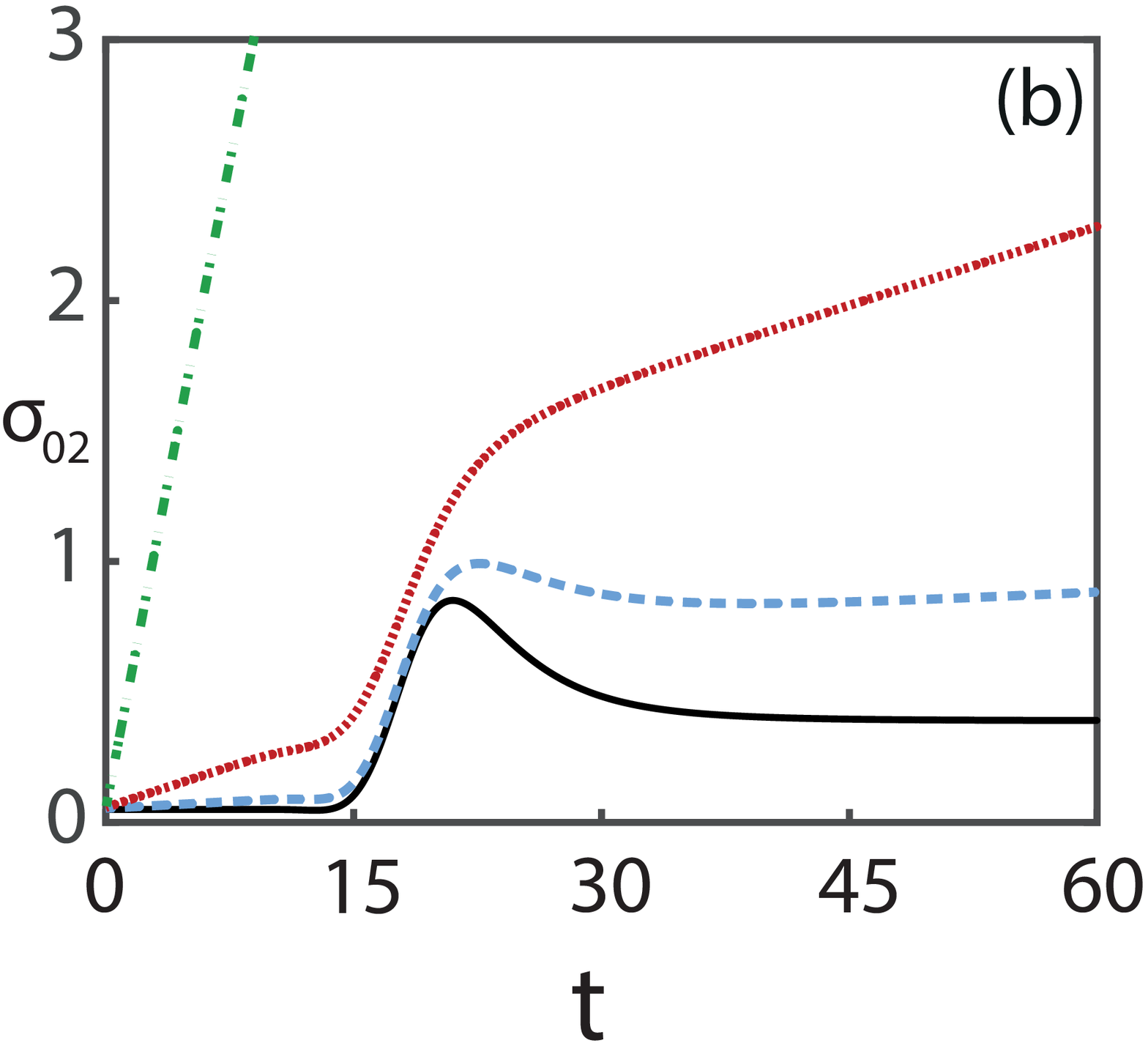,scale=0.24} 
\caption{Average momentum $<p>$ (a) and momentum spread $\sigma_{02}$ (b) in dimensionless units as functions of time for $E_K=0.5$, $\tau=3$ and $\sigma_0=0.1E_K$. The values of $\lambda$ are the same as in Fig. \ref{fig:dens05}. }\label{fig:pav05}
\end{center}
\end{figure}

\begin{figure}[ht!]
\setlength{\unitlength}{1cm}
\begin{center}
\epsfig{file=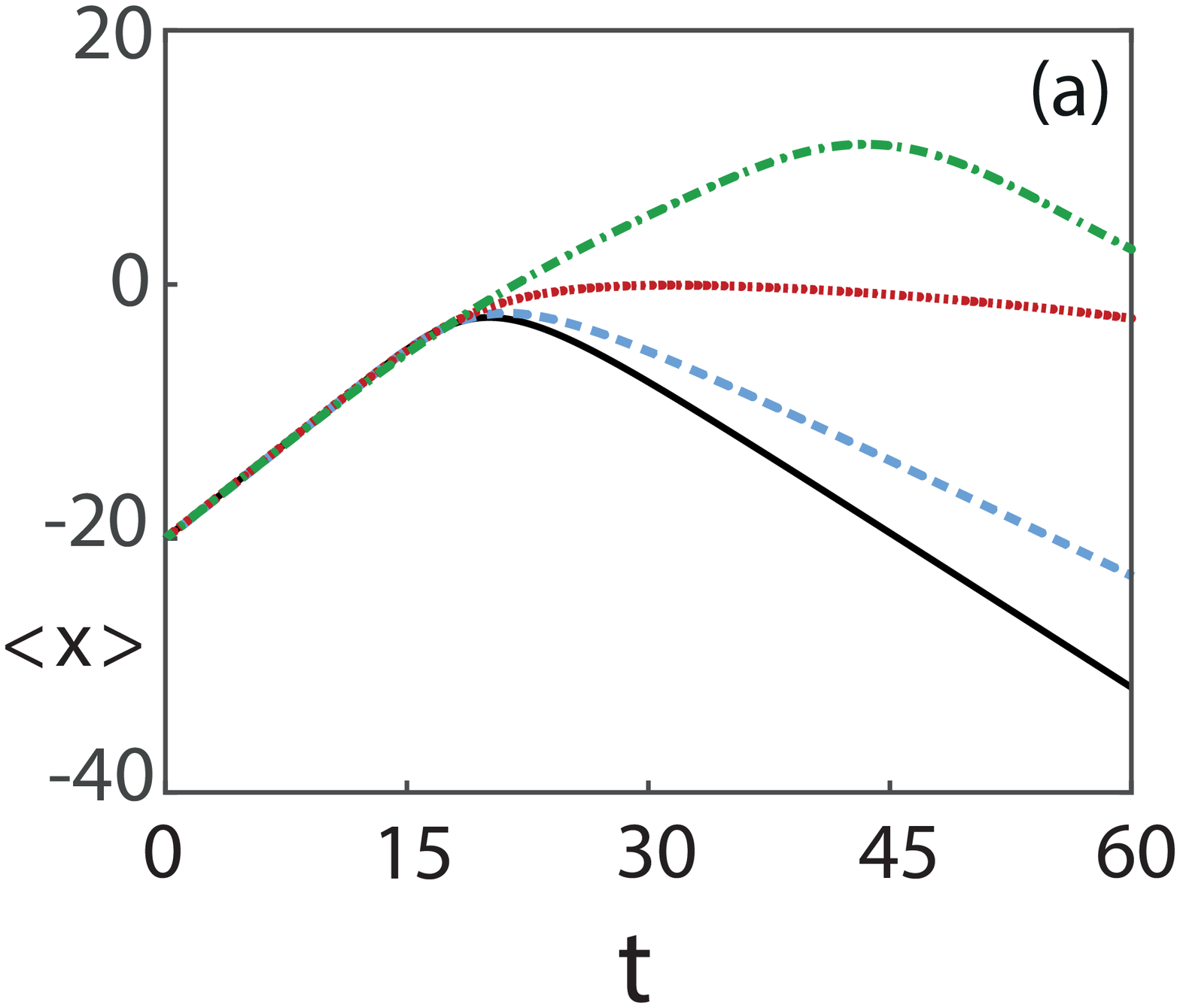,scale=0.24}\hspace*{0.25 cm}\epsfig{file=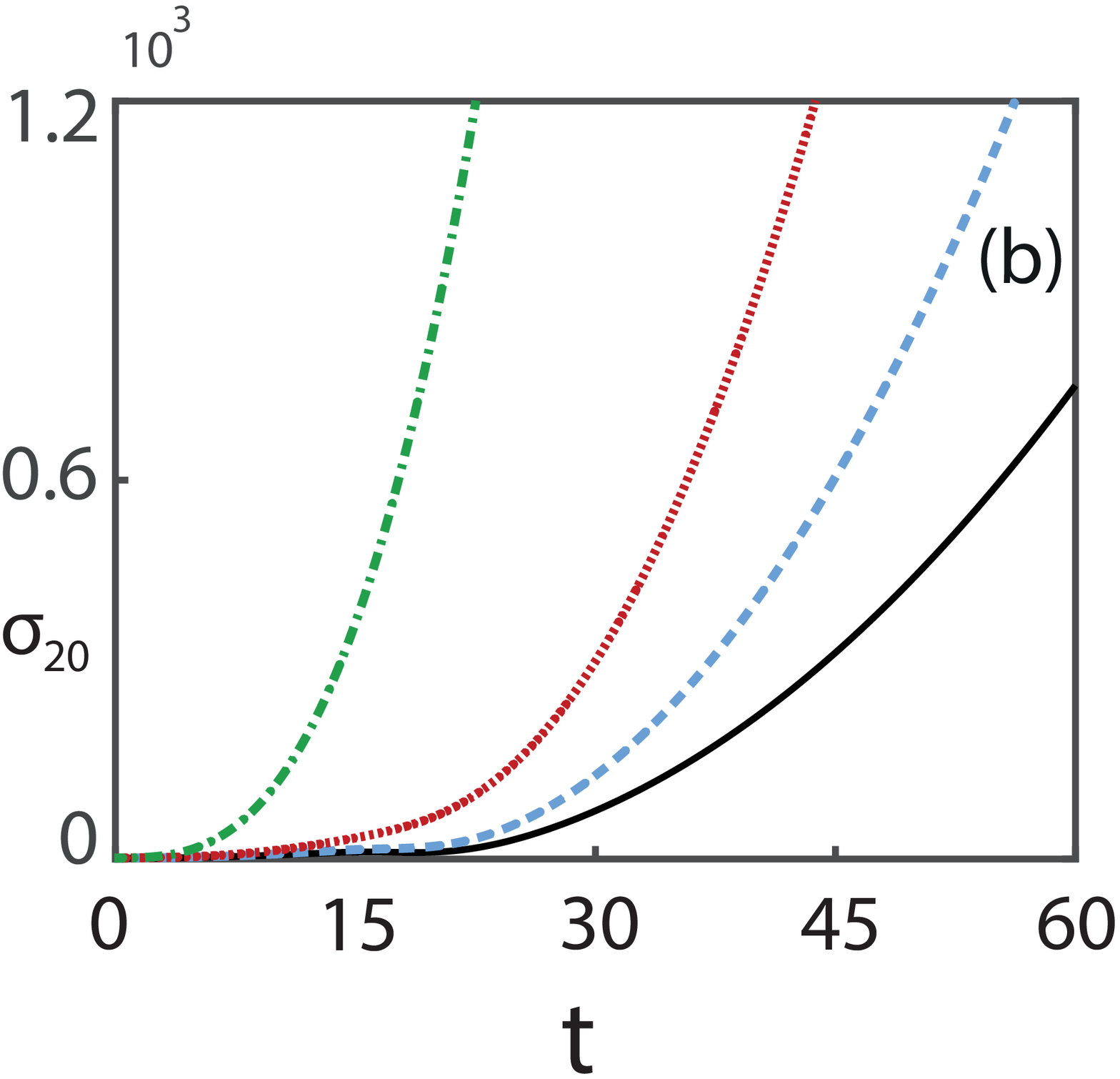,scale=0.24} 
\caption{Average position $<x>$ (a) and position spread $\sigma_{20}$ (b) in dimensionless units as functions of time for $E_K=0.5$, $\tau=3$ and $\sigma_0=0.1E_K$. The values of $\lambda$ are the same as in Fig. \ref{fig:dens05}. }\label{fig:xav05}
\end{center}
\end{figure}

\begin{figure*}[ht!]
\begin{center}
\epsfig{file=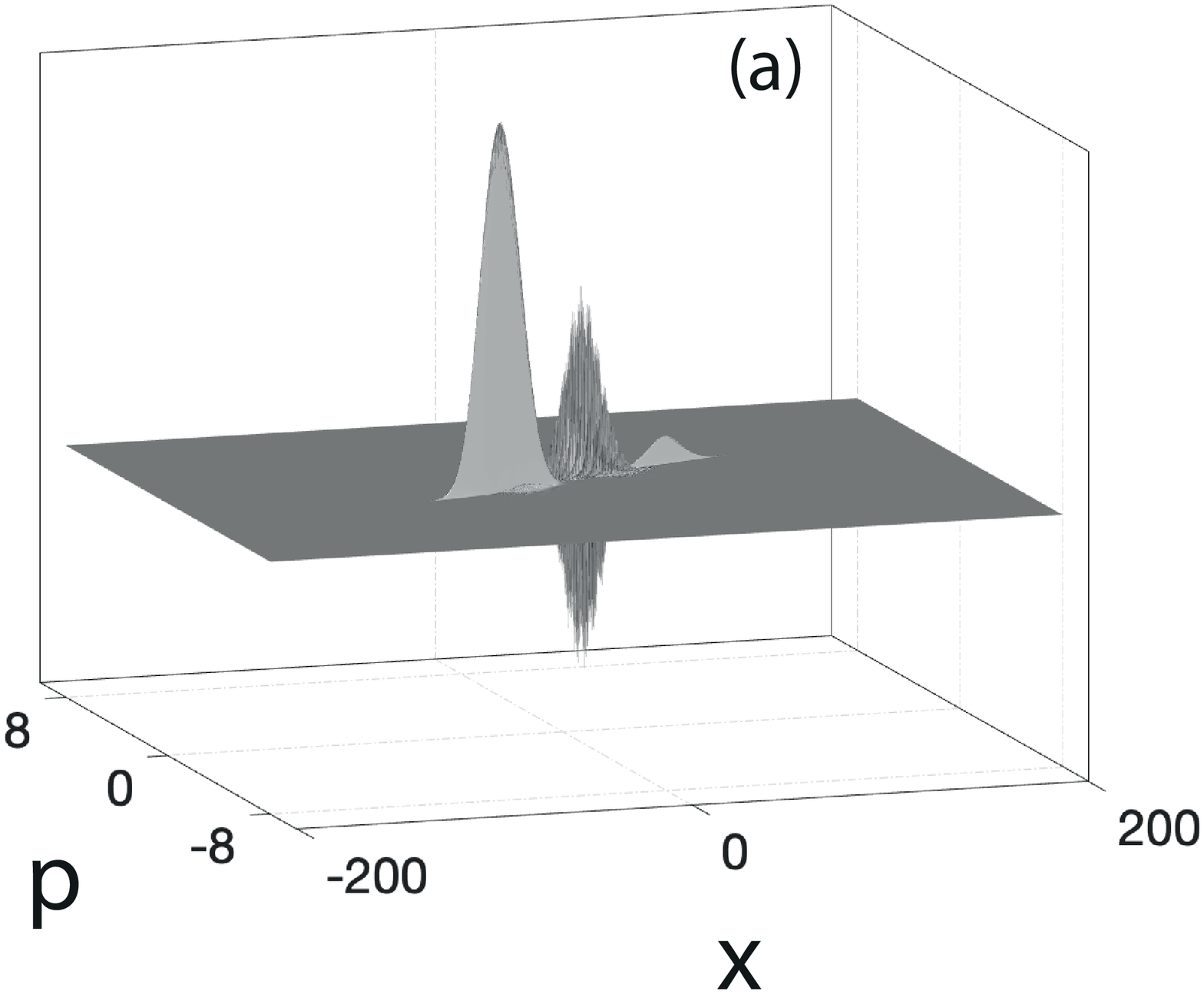,scale=0.26}\epsfig{file=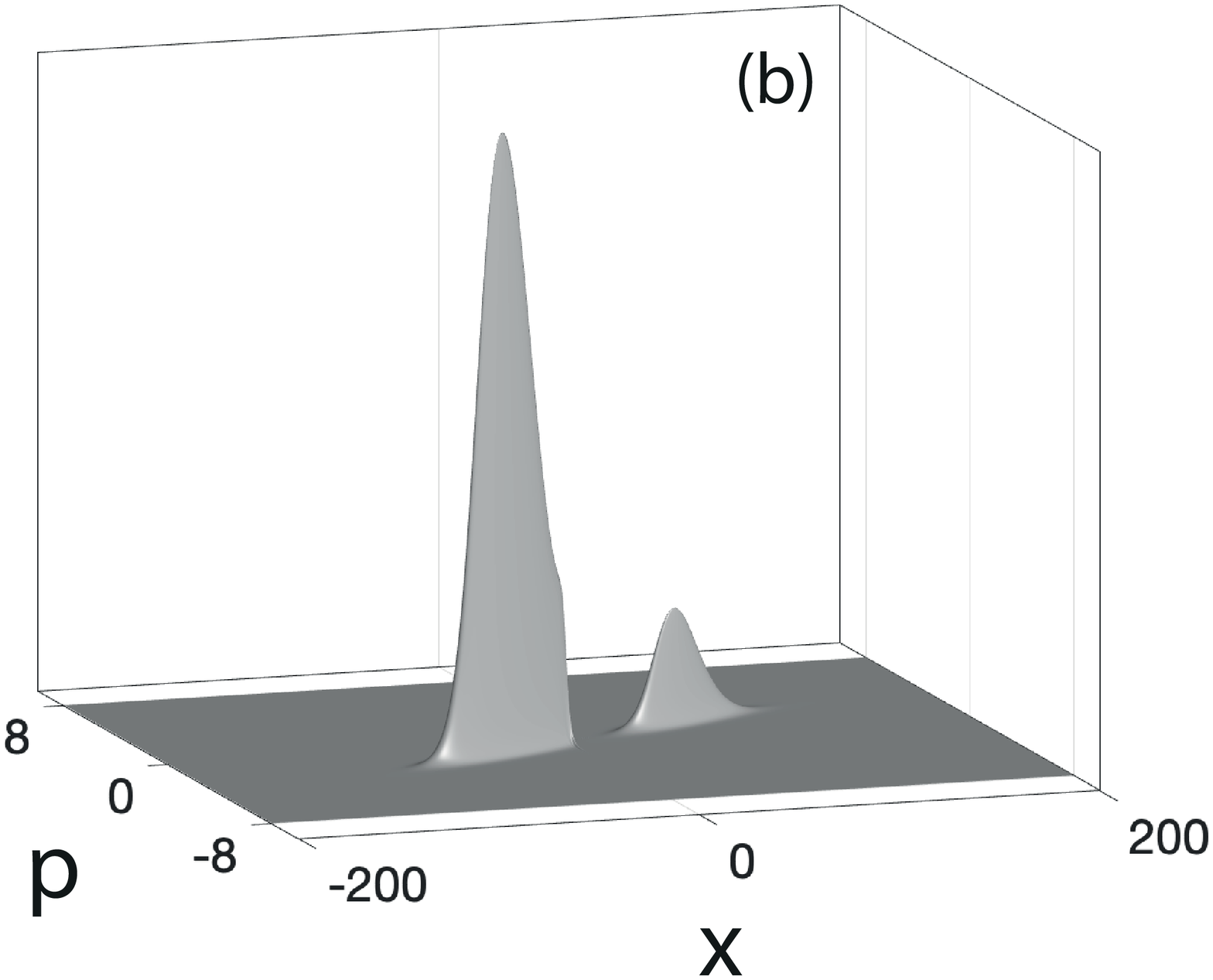,scale=0.25} 
\epsfig{file=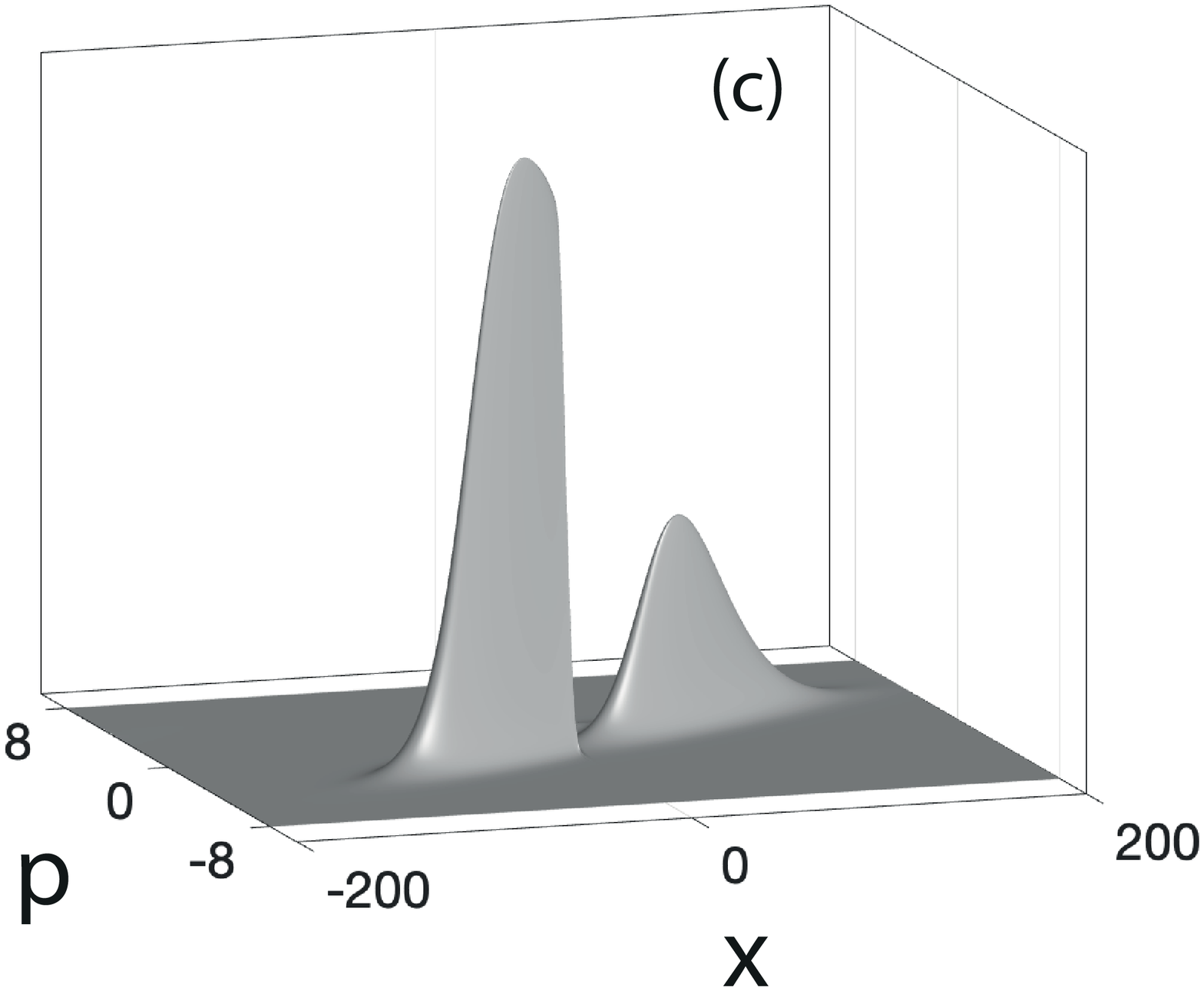,scale=0.26}\caption{Wigner function $f(x,p)$ at $t=60$, for $E_K=0.5$, $\tau=3$, $\sigma_0=0.1E_K$ and three different values of $\lambda$. Coherent case ($\lambda\rightarrow \infty$) (a); $\lambda=10$ (b); $\lambda=4$ (c).} \label{fig:wig05}
\end{center}
\end{figure*}

The dynamics of the scattering process for $E_K=0.5$ is shown in Figs. \ref{fig:dens05}-\ref{fig:xav05}, where the density profile and the average quantities are portrayed. In Figs. \ref{fig:dens05}(a)-\ref{fig:dens05}(d) we show the density profile $n(x)$ at $t=12$ (early stages of the evolution), $t=36$ (beginning of the scattering process and onset of the density oscillations), $t=48$ (oscillations begin to disappear) and $t=60$ (past the scattering process), for $\lambda=10$ (dashed blue lines), $\lambda=4$ (dotted red lines), and $\lambda=1$ (dash-dotted green lines); the standard quantum case (corresponding to $\lambda \to \infty$) is also shown for reference (solid black lines). In Figs. \ref{fig:pav05}(a) and \ref{fig:pav05}(b) the average momentum and the momentum spread, respectively,  are shown for $t \in [0,60]$ and  in \ref{fig:xav05}(a) and \ref{fig:xav05}(b) we show the average position and the position spread for the same values of $\lambda$. In Figs. \ref{fig:wig05}(a)-\ref{fig:wig05}(c) we show the Wigner function $f(x,t)$ at $t=60$ for $\lambda \to \infty$ (a), $\lambda=10$ (b), and $\lambda=4$ (c). \\

The initial Gaussian packet travels freely in the early stages of the evolution, and moves towards the potential region, with the average momentum staying constant before the bulk of the packet reaches the potential. The
packet presents the natural increasing spread, both in position and in momentum, this effect becoming more pronounced
at smaller correlation lengths [see Figs. \ref{fig:pav05}(b) and \ref{fig:xav05}(b)]. As the packet reaches the potential, the density profile exhibits oscillations, which appear to be smoothened as the correlation length becomes shorter; the average momentum, initially positive, drops and turns negative in the course of the scattering event, the drop being also less pronounced at shorter values of $\lambda$ (see Fig. \ref{fig:pav05}). \\

By looking at the density profiles, here and in the other two cases, the dependence of $n$ on $\lambda$ might seem in contradiction with Eqs. (\ref{densf}) and (\ref{densflam}), i.e.\ with the fact that $f$ and $f_\lambda$ have the same $n$.
However, the different curves corresponding to different values of $\lambda$ are the densities of {\em different} Wigner functions: they all start with the same initial datum but  they have distinct dynamics, because of the different values of $\lambda$ in the evolution equation \eqref{WED2}.

After the interaction with the potential has occurred, the transmitted and the reflected packet separate, traveling away from the origin in opposite directions, the reflected packet being considerably larger than the transmitted one in the standard quantum case. As the correlation length is made smaller, the Wigner function spreads out more both in momentum and in position, causing the transmitted portion to increase and the reflected portion  to decrease in size, thus making the decrease of the average momentum less pronounced. In the standard quantum case, in addition to the reflected and the transmitted portions, the Wigner function displays a strongly oscillating behavior near the potential region [see Fig. \ref{fig:wig05}(a)]; as the correlation length becomes smaller, the oscillations are seen to be damped away [see Figs. \ref{fig:wig05}(b) and \ref{fig:wig05}(c)]. \\

In addition to the damping of the oscillations, two  important additional effects of the finite correlation length on  the dynamics of the Wigner function must be pointed out. \\

\noindent {\it (i)} First of all, the broadening and flattening of the Wigner function with the decreased correlation length can be observed both in time [for fixed $\lambda$, $f(x,p,t)$ becomes  flatter and broader as $t \to \infty$] and with respect to $\lambda$ [for fixed $t$, $f(x,p,t)$ becomes  flatter and broader as $\lambda \to 0$]. This flattening and broadening effect of the correlation length is in agreement with the results of \cite{Barletti18,Adami03,Joos85}, where the increased transmission due to the decoherence was attributed to a reduced momentum exchanged between the packet and the potential caused by the correlation damping. As a consequence, this poses a difficulty in the description of the long-time evolution of the system by the Wigner-function model with finite correlation length; it also leads to numerical problems, because of the need for a larger and larger simulation domain as 
$t \to \infty$ and $\lambda \to 0$.  These theoretical and numerical difficulties are well represented by the dash-dotted green curves in Figs. 
\ref{fig:dens05}(a)-\ref{fig:xav05}(b), corresponding to $\lambda=1$: the curves showing the average quantities behave unphysically and the very flat density profile seen in Fig. \ref{fig:dens05}(d) strongly suggests that the simulation domain should be made larger. In the next examples at higher energies we will not show any longer the results for $\lambda=1$ at the later times. The difficulty in applying the decoherence model to the large-time behavior was already briefly mentioned in Sec. \ref{sec:WignerDecoherence}

\noindent {\it (ii)} Another feature which is observed in the particle density is the sharp jump in the profile at $x=0$ (see Figs. \ref{fig:dens05}(a)-\ref{fig:dens05}(d)), the jump being larger in relative magnitude (with respect to the peak of the density) as $\lambda$ decreases. The jump in the density profile is directly related to the apparent discontinuity seen in the Wigner function at $x=0$ [see Figs. \ref{fig:wig05}(b) and \ref{fig:wig05}(c)] for $\lambda=10$ and $\lambda=4$; here, the reflected and the transmitted packets appear well separated in the potential region, where the Wigner function takes very small values. An analysis and an explanation of this phenomenon will be developed in Sec. \ref{sub:Discontinuity}.

%\clearpage
%
\subsection{Second case, intermediate regime: $E_K=1$}
\begin{figure}[ht!]
\begin{center}
\epsfig{file=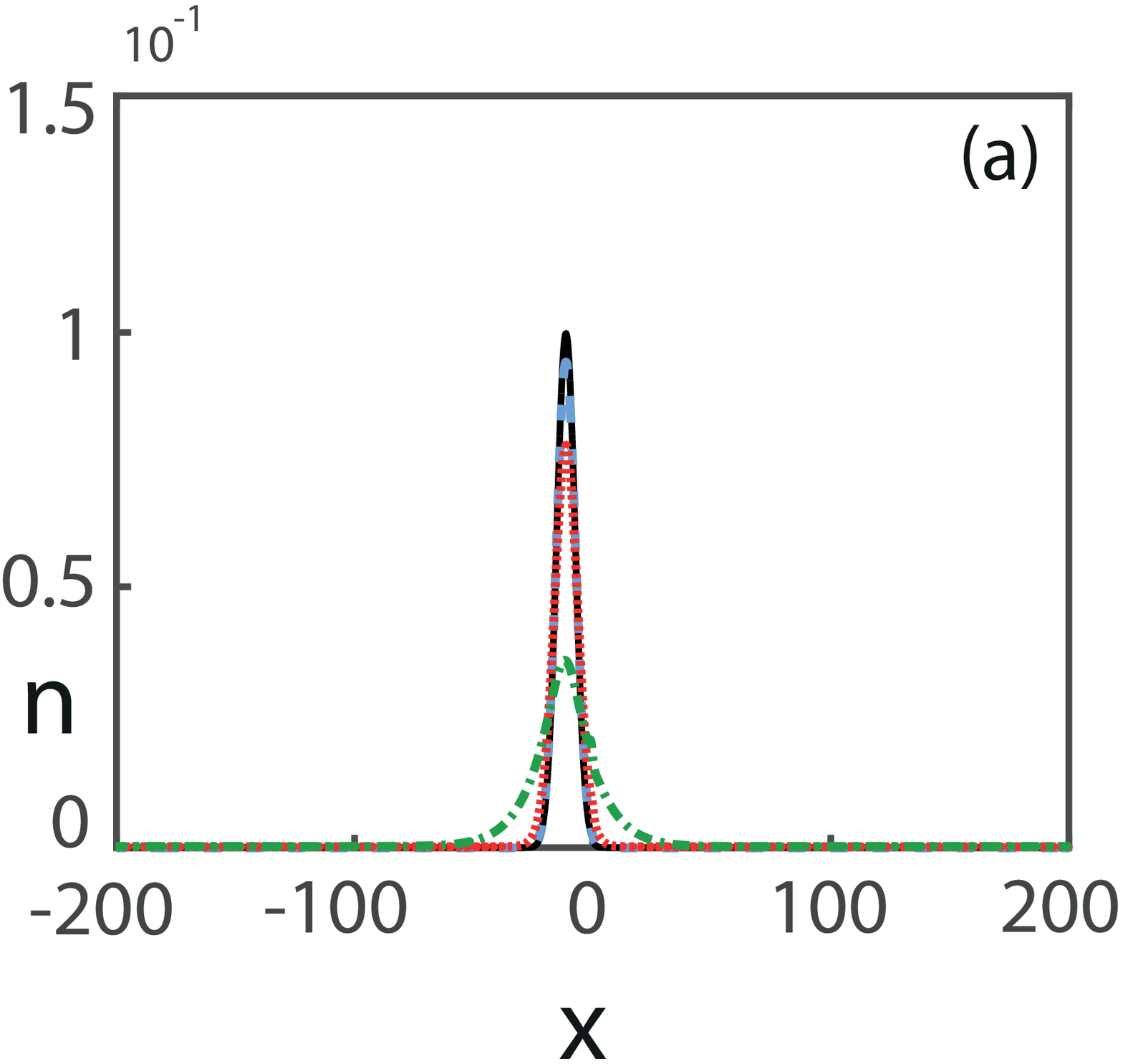,scale=0.24}\hspace*{0.25 cm}\epsfig{file=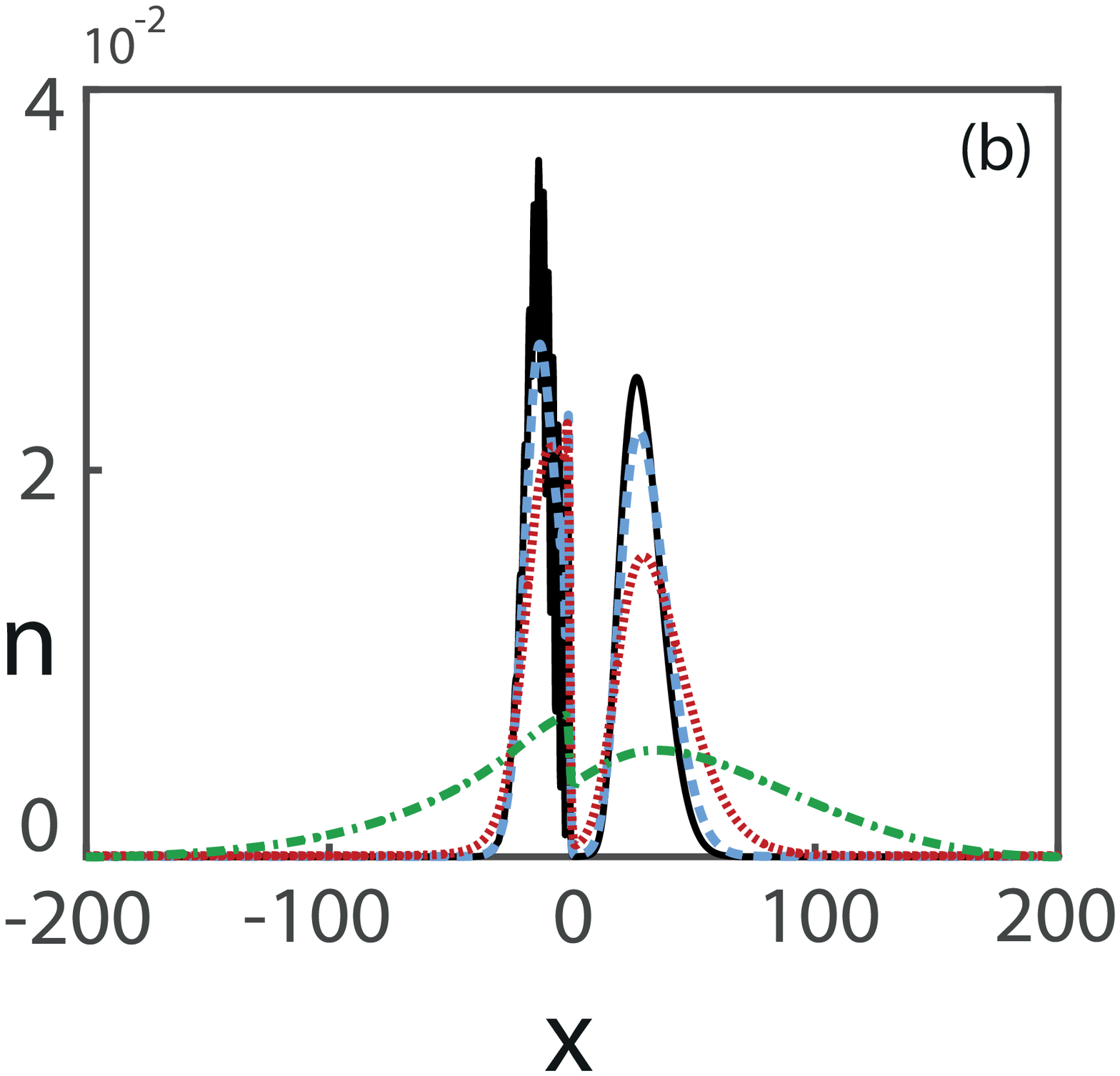,scale=0.24} 
\epsfig{file=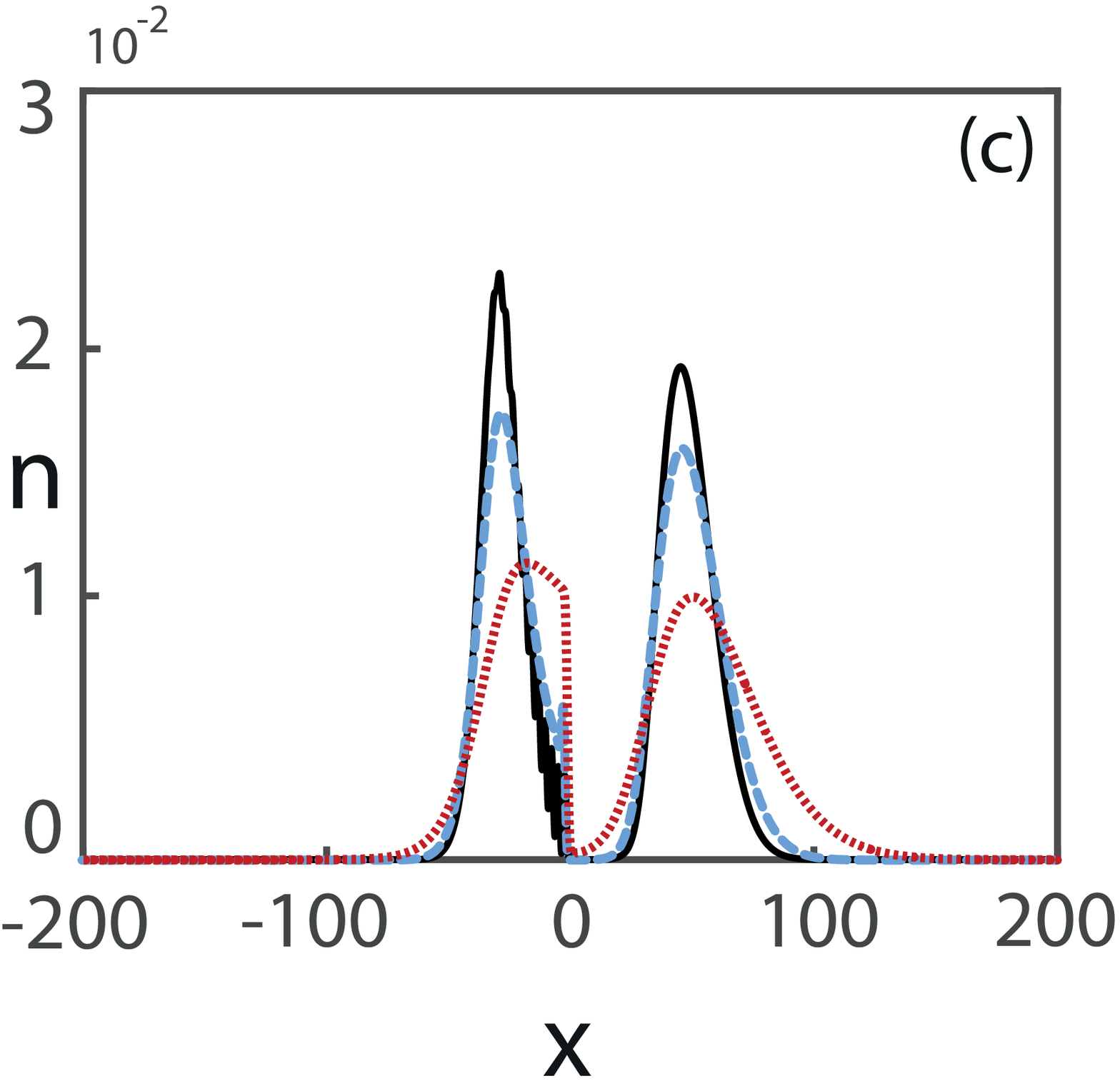,scale=0.24}\hspace*{0.25 cm}\epsfig{file=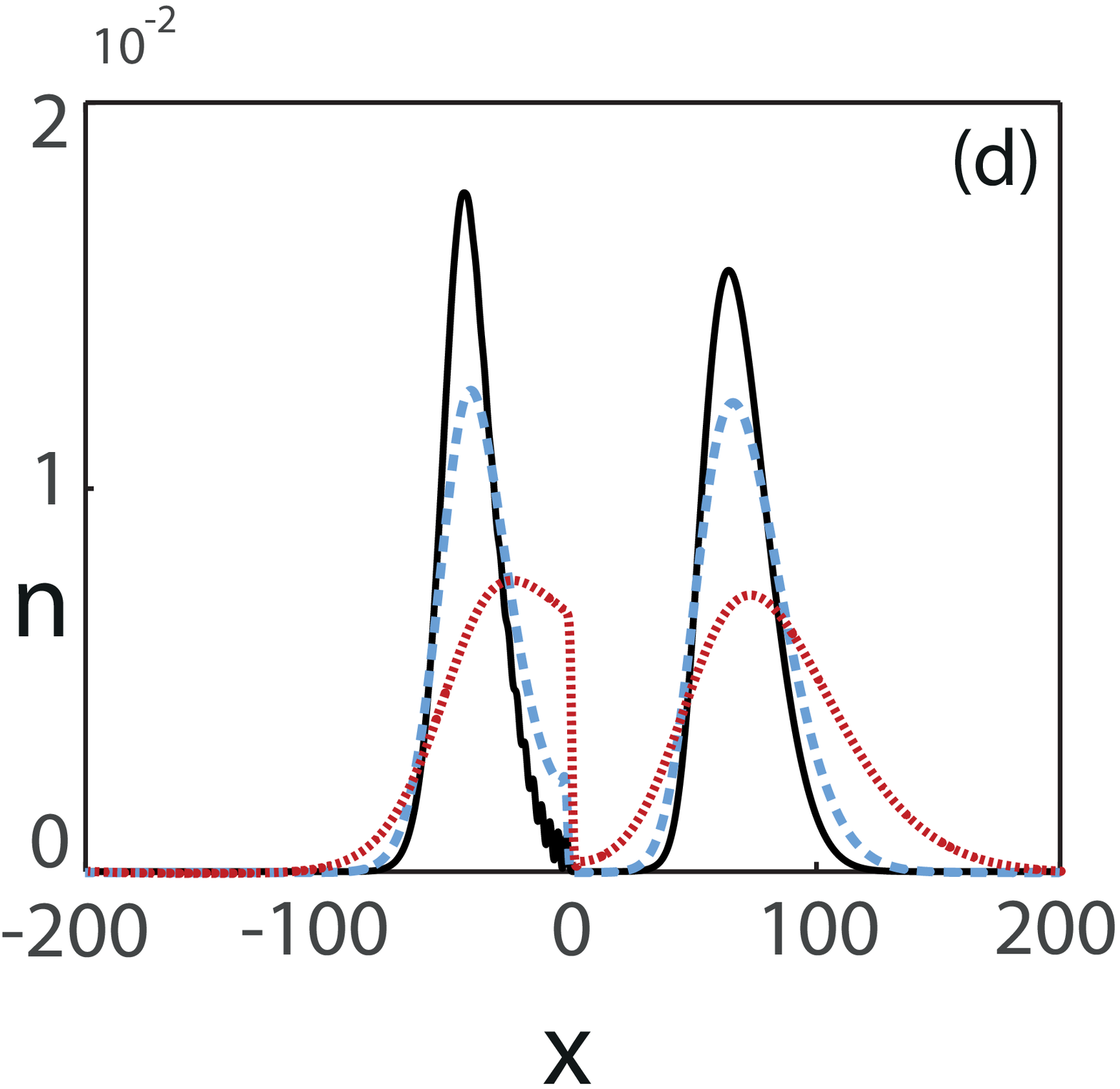,scale=0.24} 
\caption{Density profile $n(x)$ in dimensionless units as a function of $x$ at four instants of time, $t=12$ (a), $t=36$ (b), $t=48$ (c), and $t=60$ (d) for $E_K=1$, $\tau=3$ and $\sigma_0=0.1E_K$. The values of $\lambda$ are the same as in Fig. \ref{fig:dens05} ($\lambda=1$ (dash-dotted green line) in Figs. (a) and (b) only).} \label{fig:dens1}
\end{center}
\end{figure}
\begin{figure}[ht!]
\setlength{\unitlength}{1cm}
\begin{center}
\epsfig{file=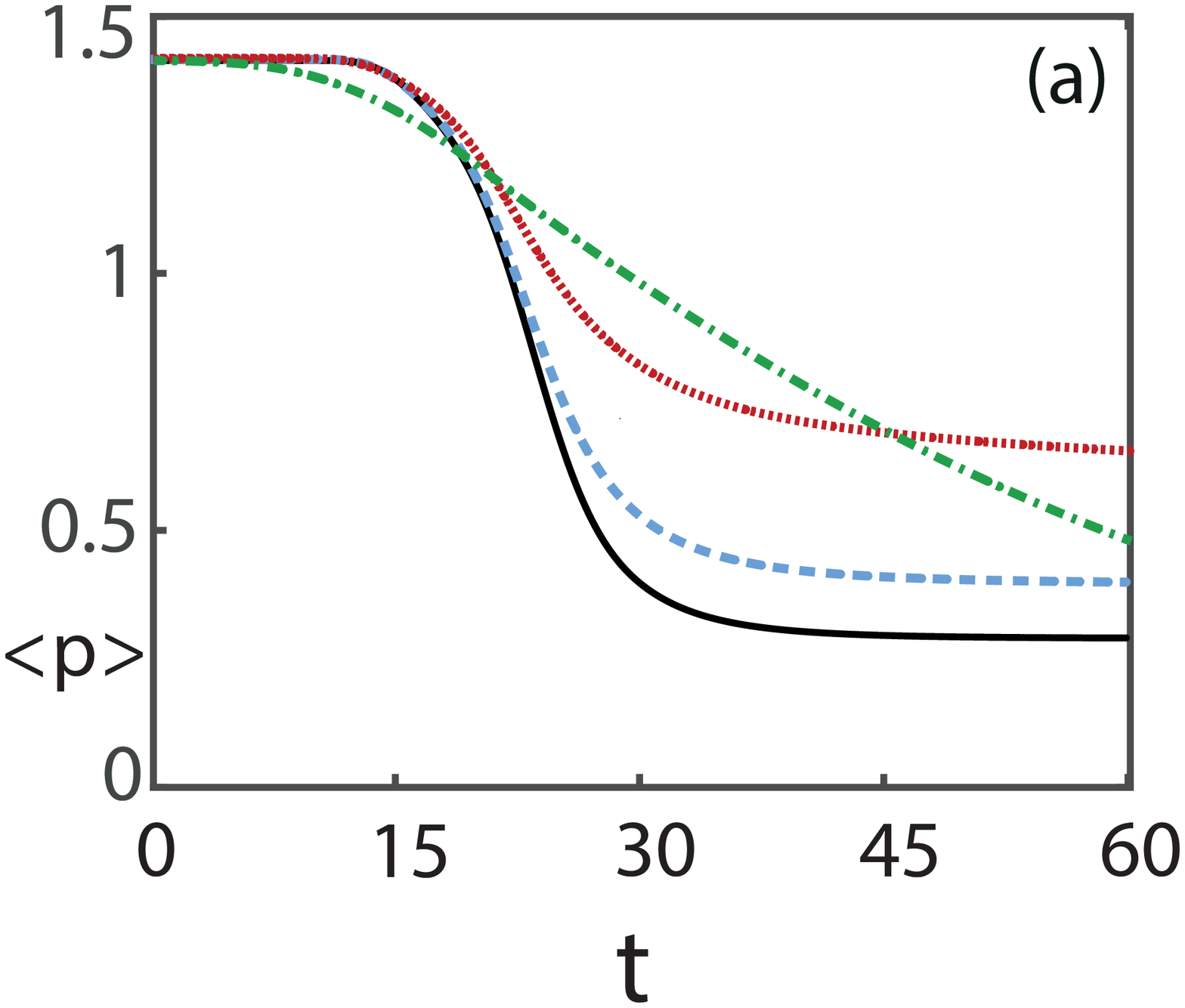,scale=0.24}\hspace*{0.1 cm}\epsfig{file=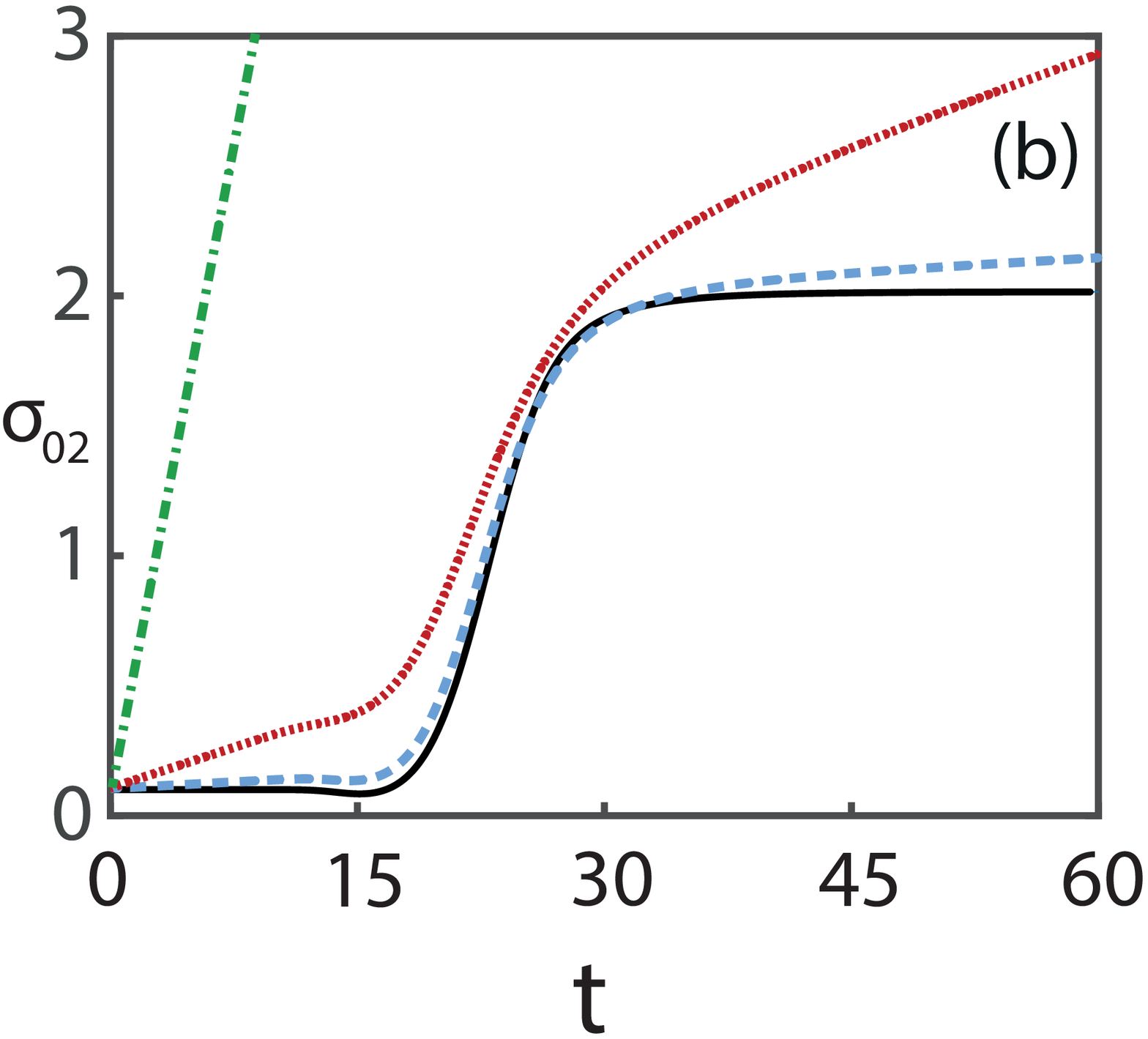,scale=0.24} 
\caption{Average momentum $<p>$ (a) and momentum spread $\sigma_{02}$ (b) in dimensionless units as functions of time for $E_K=1$, $\tau=3$ and $\sigma_0=0.1E_K$. The values of $\lambda$ are the same as in Fig. \ref{fig:dens05}. } \label{fig:pav1}
\end{center}
\end{figure}
\begin{figure}[ht!]
\setlength{\unitlength}{1cm}
\begin{center}
\epsfig{file=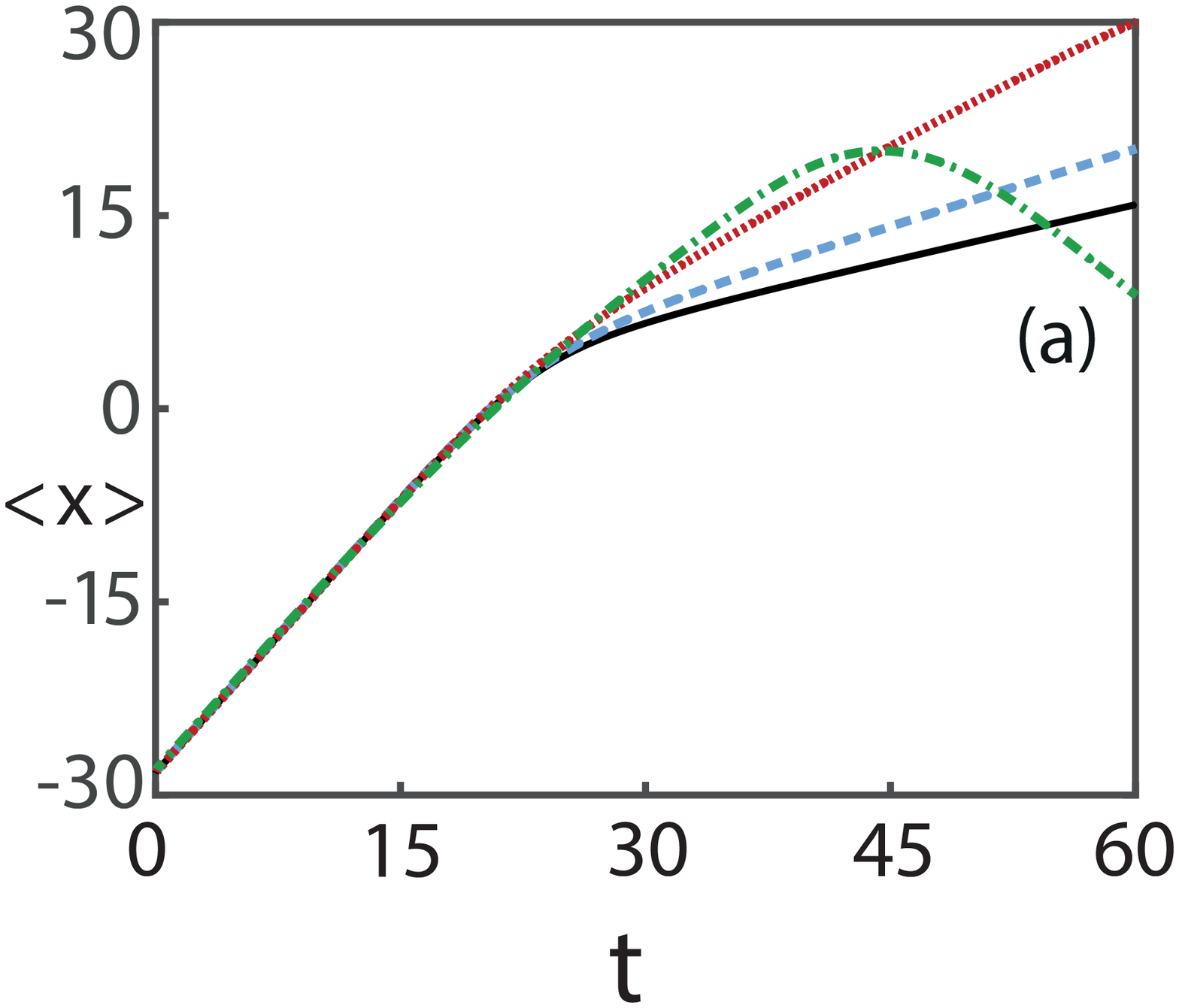,scale=0.24}\hspace*{0.1 cm}\epsfig{file=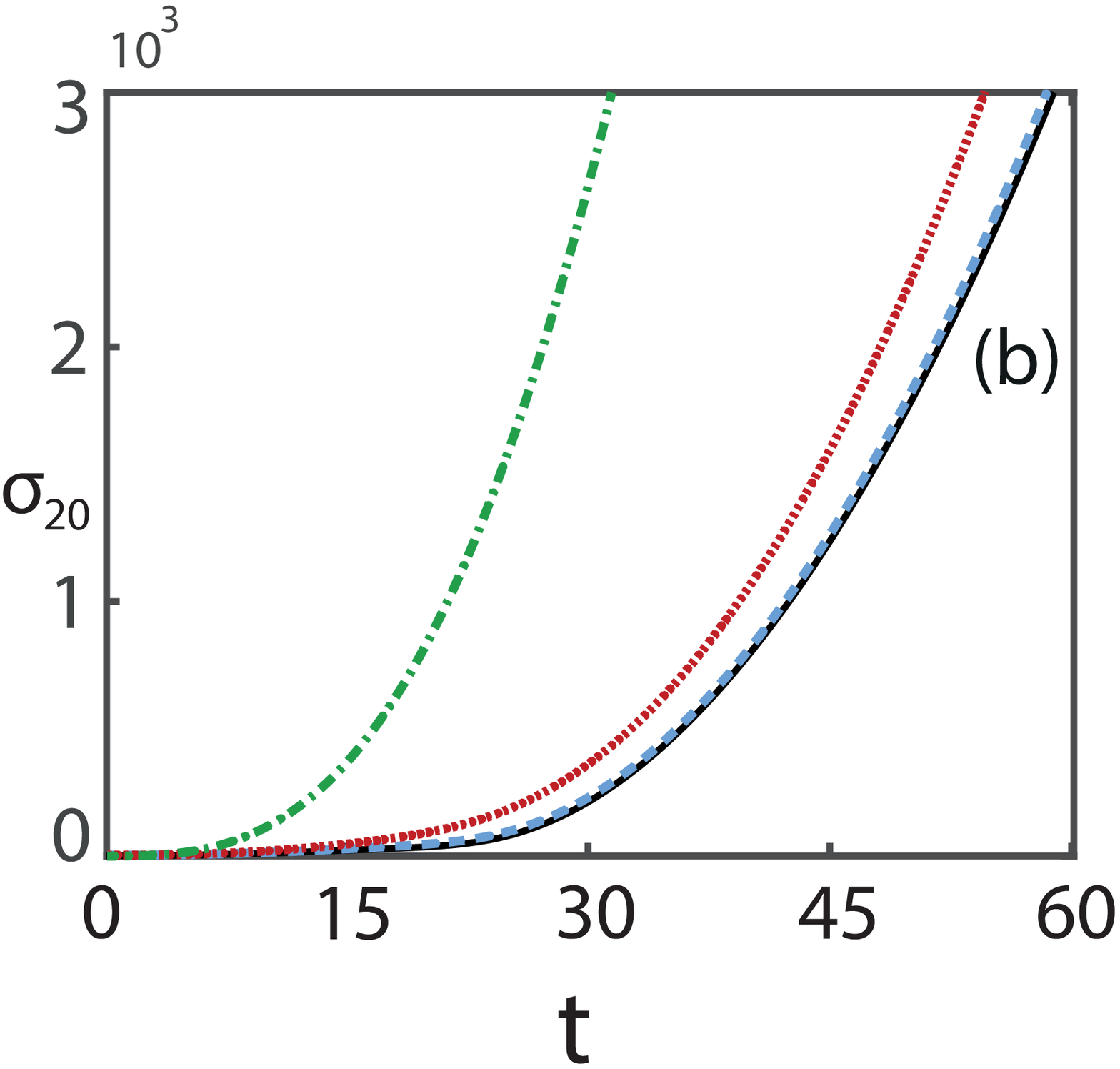,scale=0.24} 
\caption{Average position $<x>$ (a) and position spread $\sigma_{20}$ (b) in dimensionless units as functions of time for $E_K=1$, $\tau=3$, and $\sigma_0=0.1E_K$.The values of $\lambda$ are the same as in Fig. \ref{fig:dens05}. }\label{fig:xav1}
\end{center}
\end{figure}
\begin{figure*}[ht!]
\begin{center}
\epsfig{file=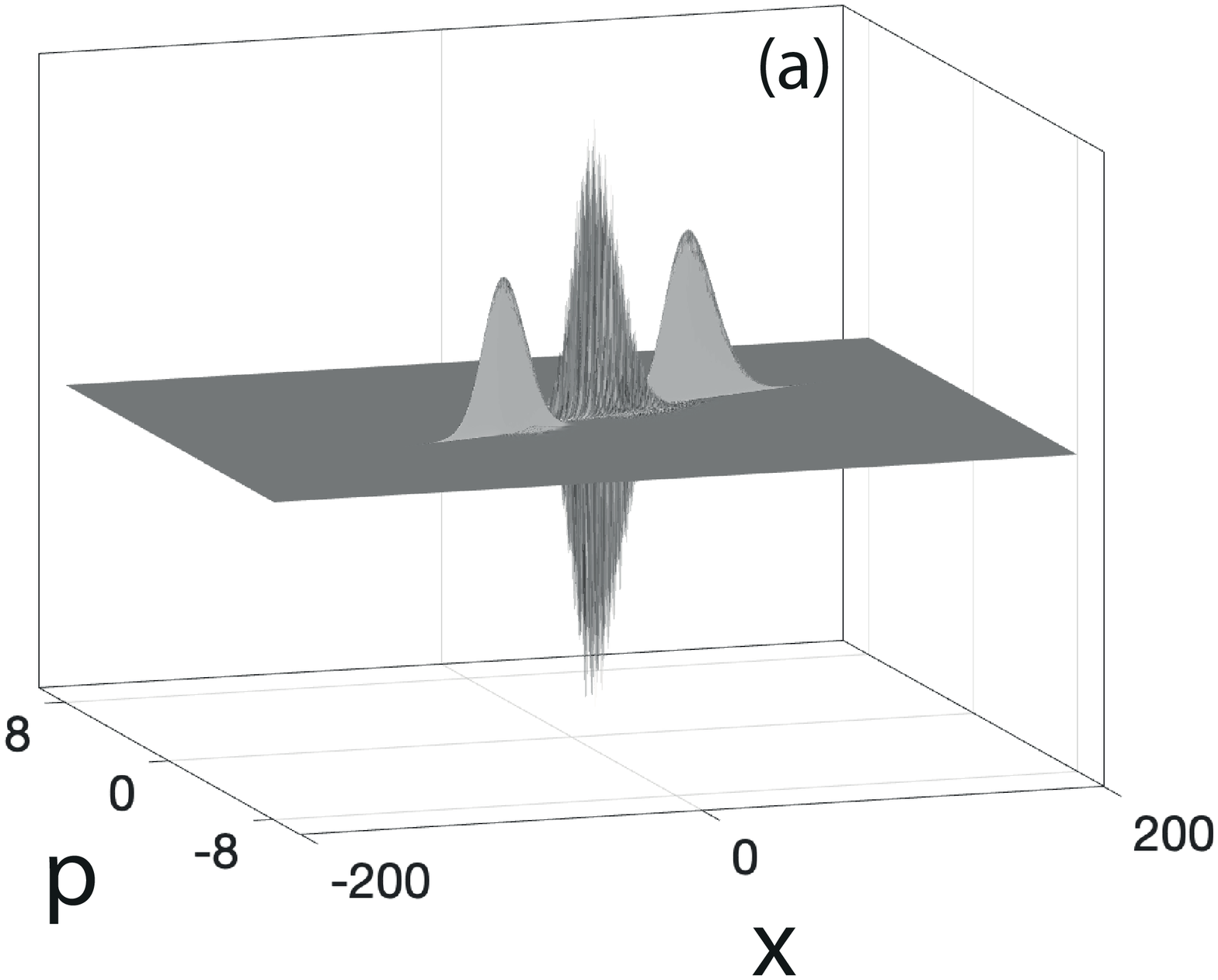,scale=0.24}\epsfig{file=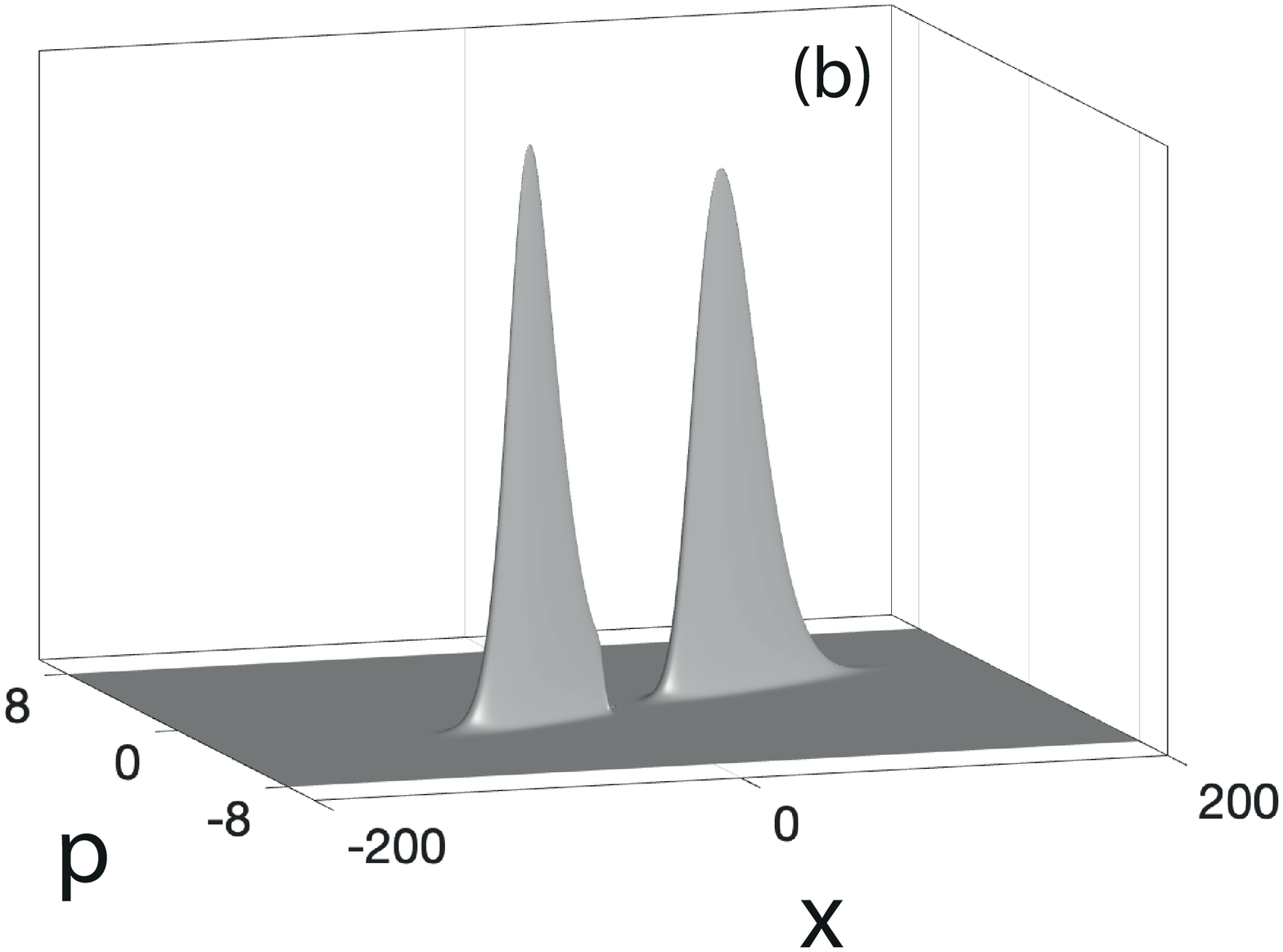,scale=0.24} 
\epsfig{file=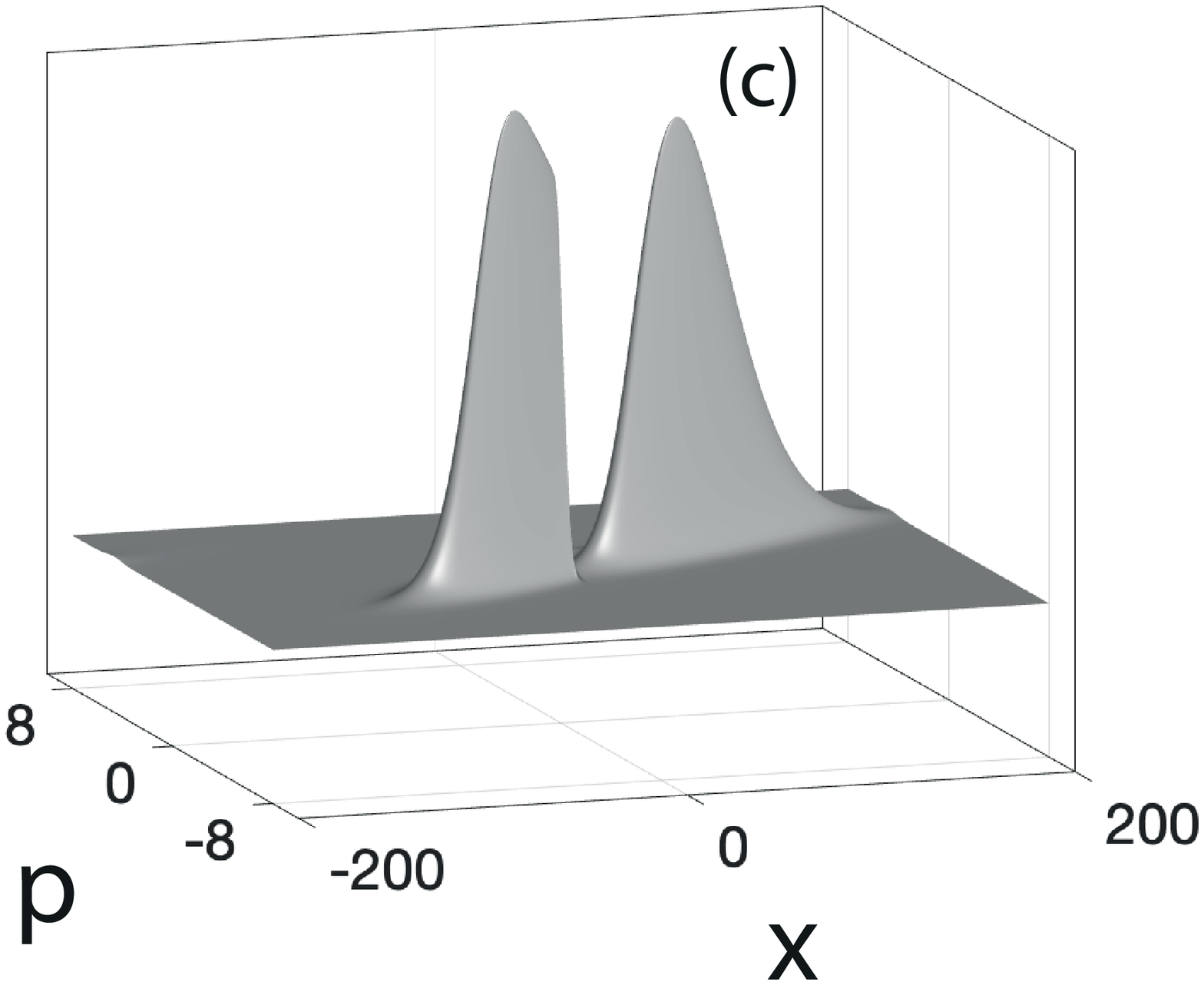,scale=0.29}
\caption{Wigner function $f(x,p)$ at $t=60$, for $E_K=1$, $\tau=3$, and $\sigma_0=0.1E_K$. The values of $\lambda$ are the same as in Fig. \ref{fig:wig05}.} \label{fig:wig1}
\end{center}
\end{figure*}

The density profiles for this case are shown in Figs. \ref{fig:dens1}(a)-\ref{fig:dens1}(d) at the same four instants of time as for the case with $E_K=0.5$, the average momentum and the momentum spread in Figs. \ref{fig:pav1}(a) and \ref{fig:pav1}(b), the average position and the position spread in Figs. \ref{fig:xav1}(a) and \ref{fig:xav1}(b), and the Wigner function in Figs. \ref{fig:wig1}(a)-\ref{fig:wig1}(c), all of them for the same values of the correlation length as for $E_K=0.5$. The initial energy is here at the same level of the potential height and the early evolution of the initial Gaussian packet is similar to the evolution observed in the previous case. 
The onset of the oscillations on the density profile is observed again during the scattering process, but a portion of the packet has already traveled past the potential region before the oscillations stop. As the interaction with the potential becomes negligible, we observe again a transmitted and a reflected packet, of approximately equal size, traveling away from the origin in opposite directions; both the transmitted and the reflected packets become broader and smaller as $\lambda$ is decreased and as $t \to \infty$. The average quantities follow the same qualitative pattern seen in the $E_K=0.5$ case: the average momentum drops after the initial constant profile, the drop being less pronounced at small values of $\lambda$. We observe again the unphysical behavior at short correlation lengths; the density jump at $x=0$ is also present.
%\clearpage
%
%
\subsection{Third case, transmission-dominated regime: $E_K=1.5$.}
\begin{figure}[ht!]
\begin{center}
\epsfig{file=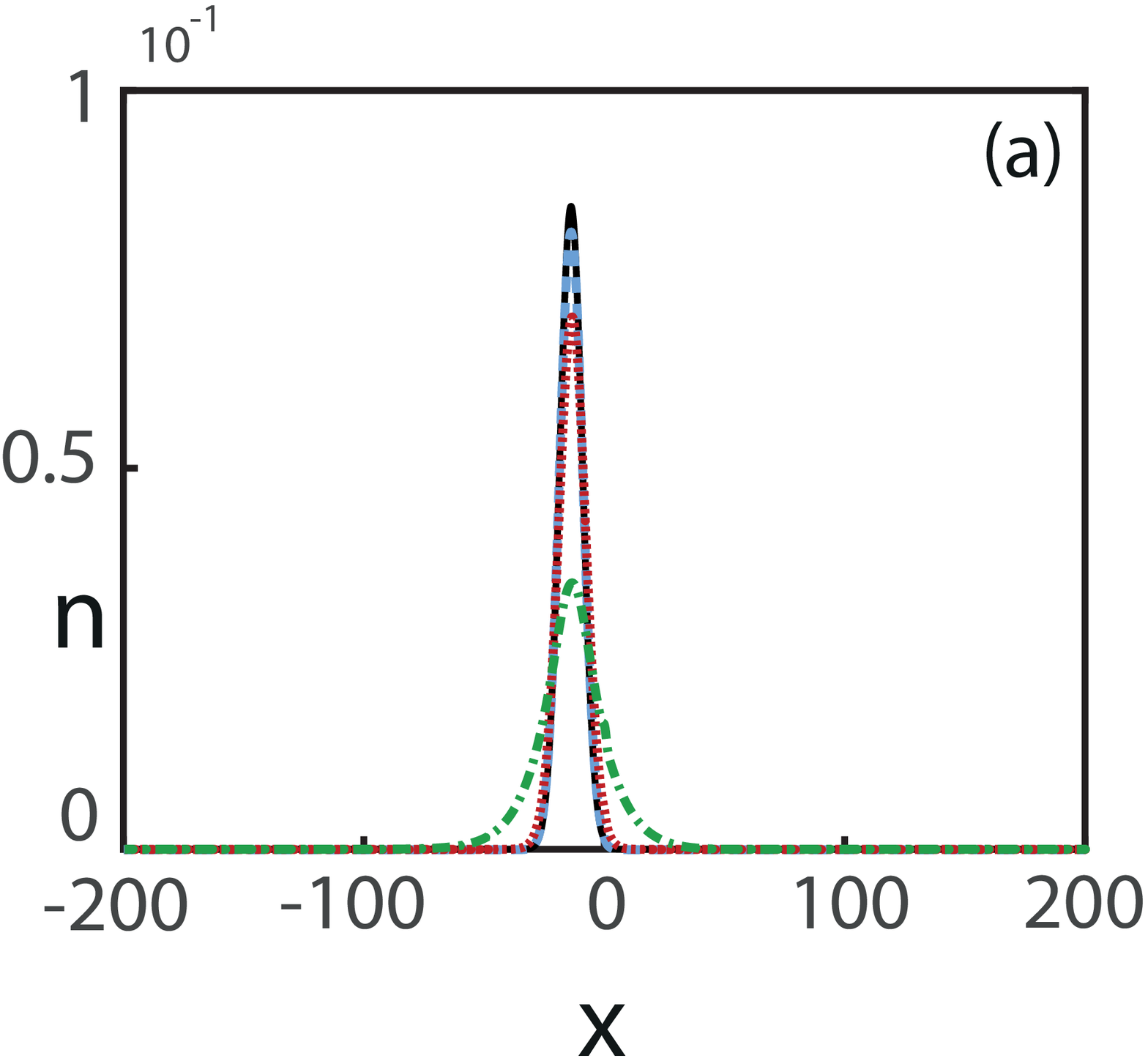,scale=0.24}\hspace*{0.25 cm}\epsfig{file=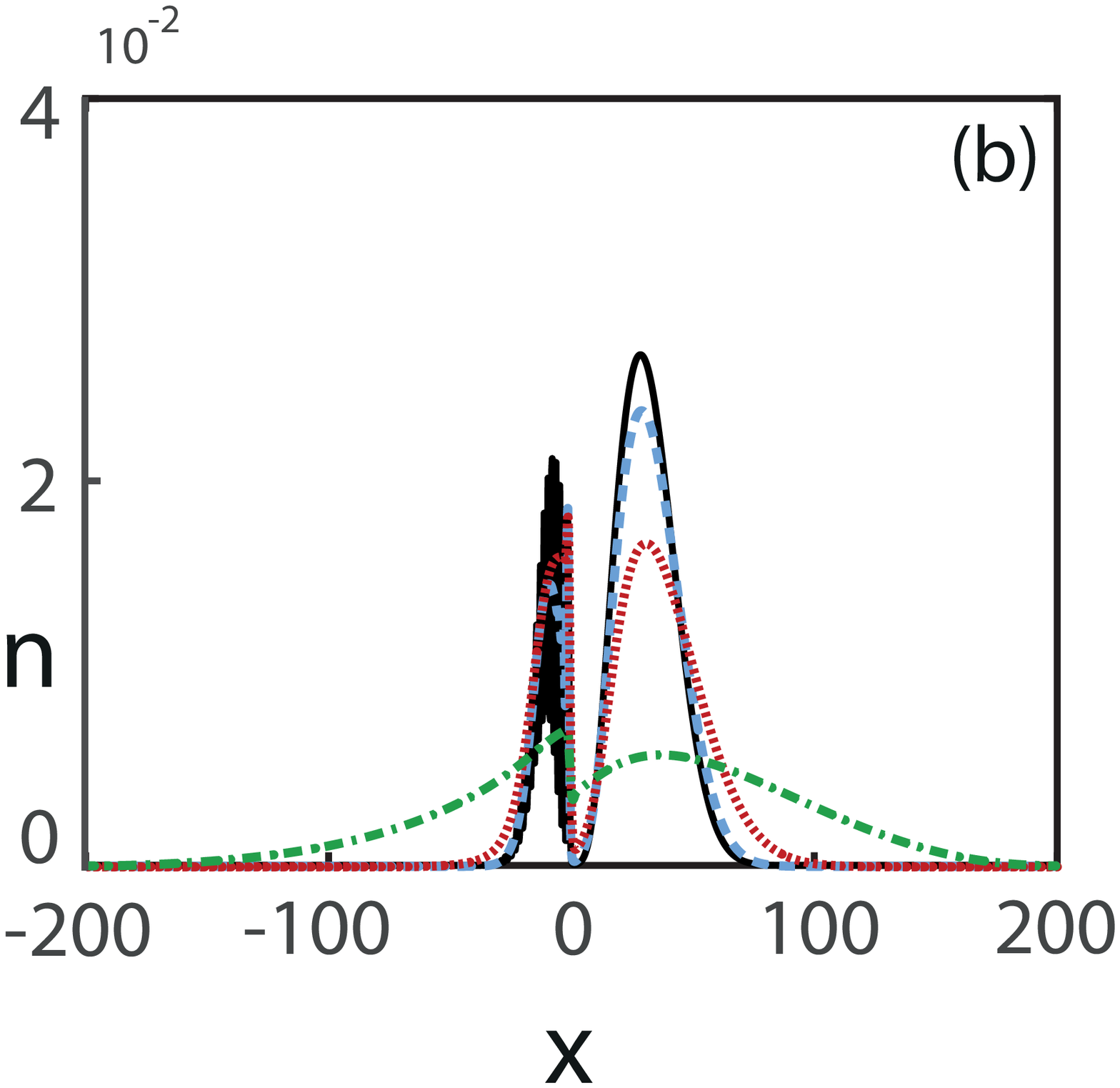,scale=0.24} 
\epsfig{file=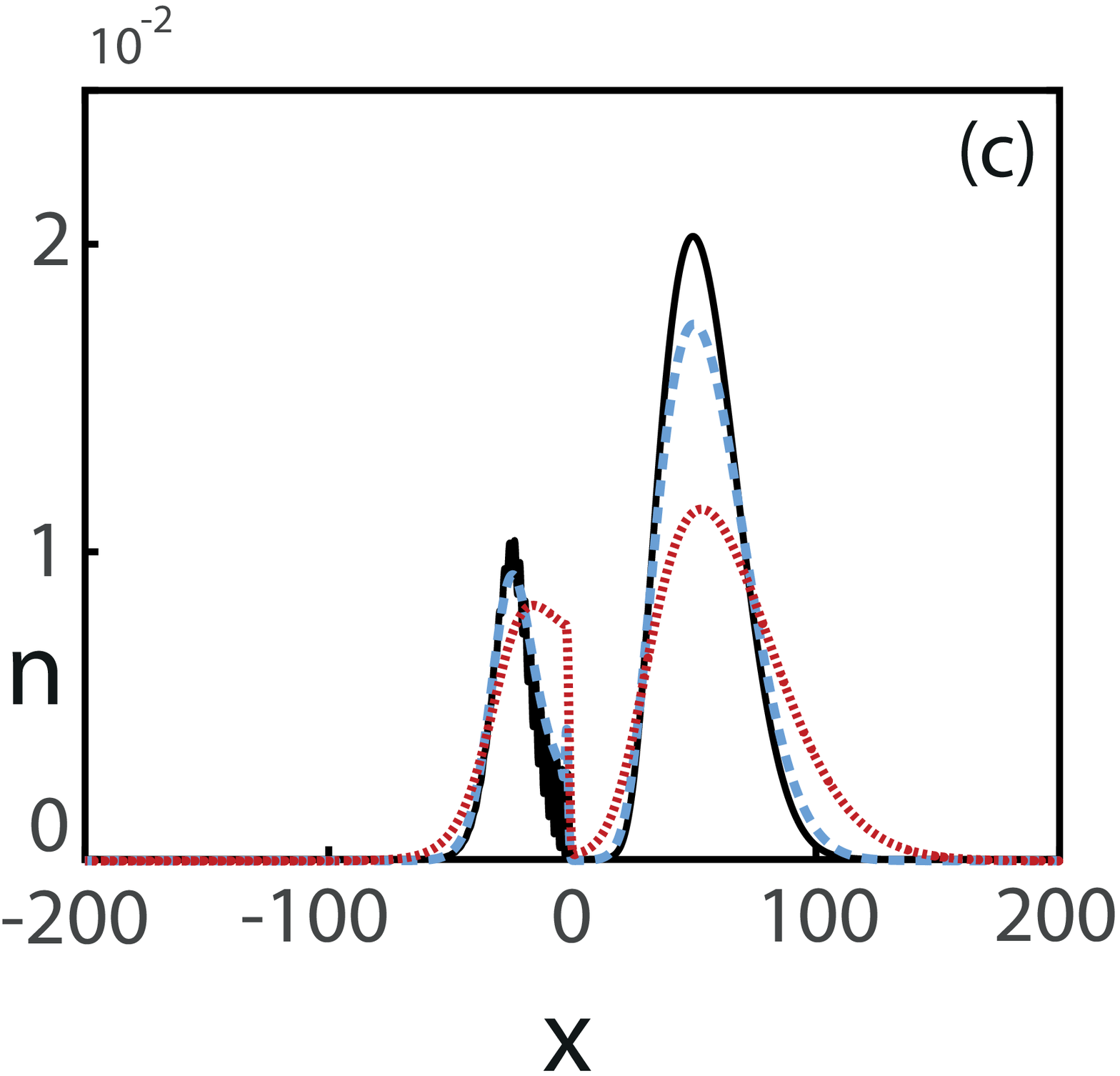,scale=0.24}\hspace*{0.25 cm}\epsfig{file=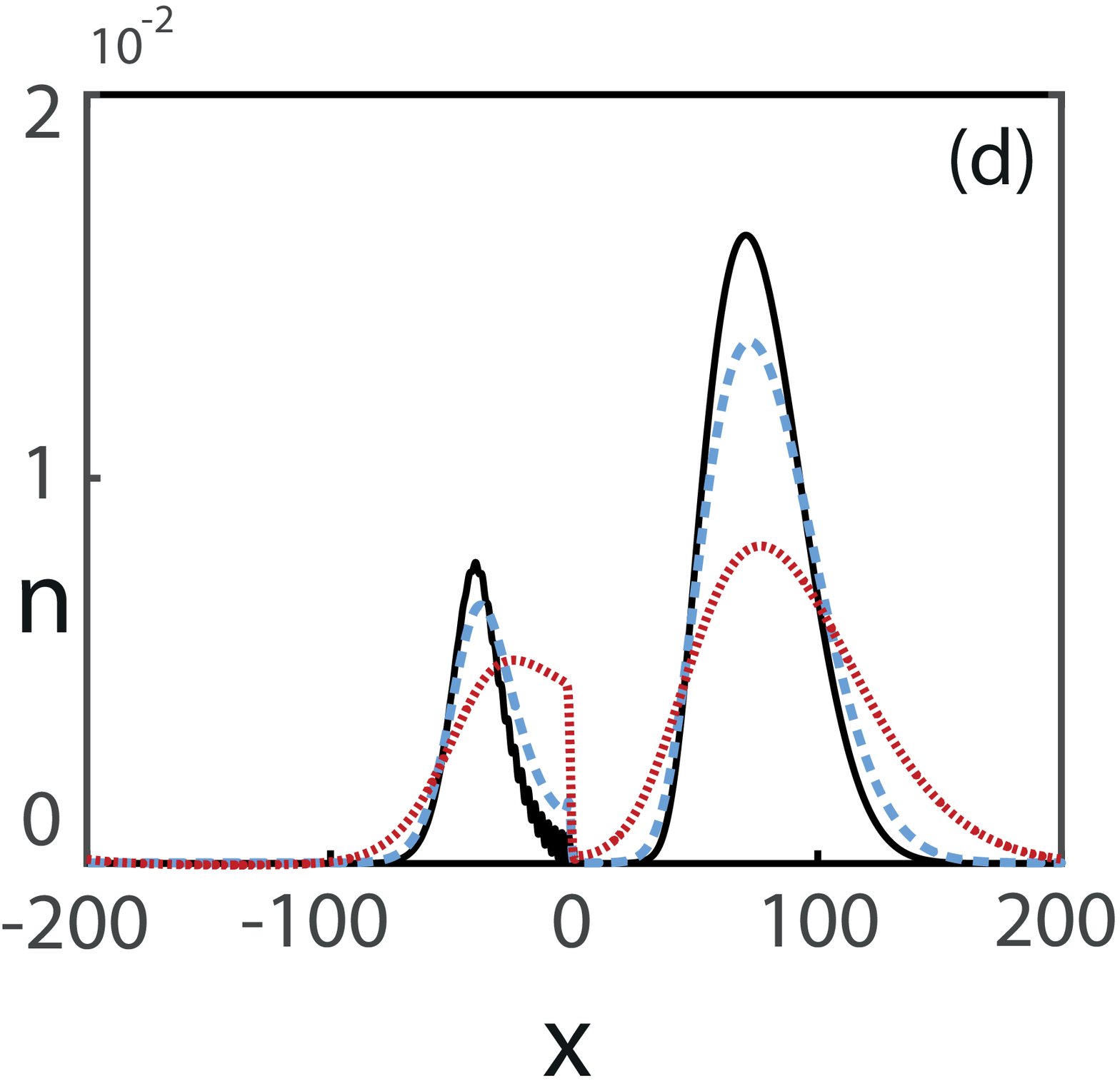,scale=0.24} 
\caption{Density profile $n(x)$ in dimensionless units as function of $x$ at four instants of time, $t=12$ (a), $t=36$ (b), $t=48$ (c), and $t=60$ (d) for $E_K=1.5$, $\tau=3$, and $\sigma_0=0.1E_K$. The values of $\lambda$ are the same as in Fig. \ref{fig:dens05} ($\lambda=1$ (dash-dotted green line) in Figs. (a) and (b) only). } \label{fig:dens15}
\end{center}
\end{figure}

\begin{figure}[ht!]
\setlength{\unitlength}{1cm}
\begin{center}
\epsfig{file=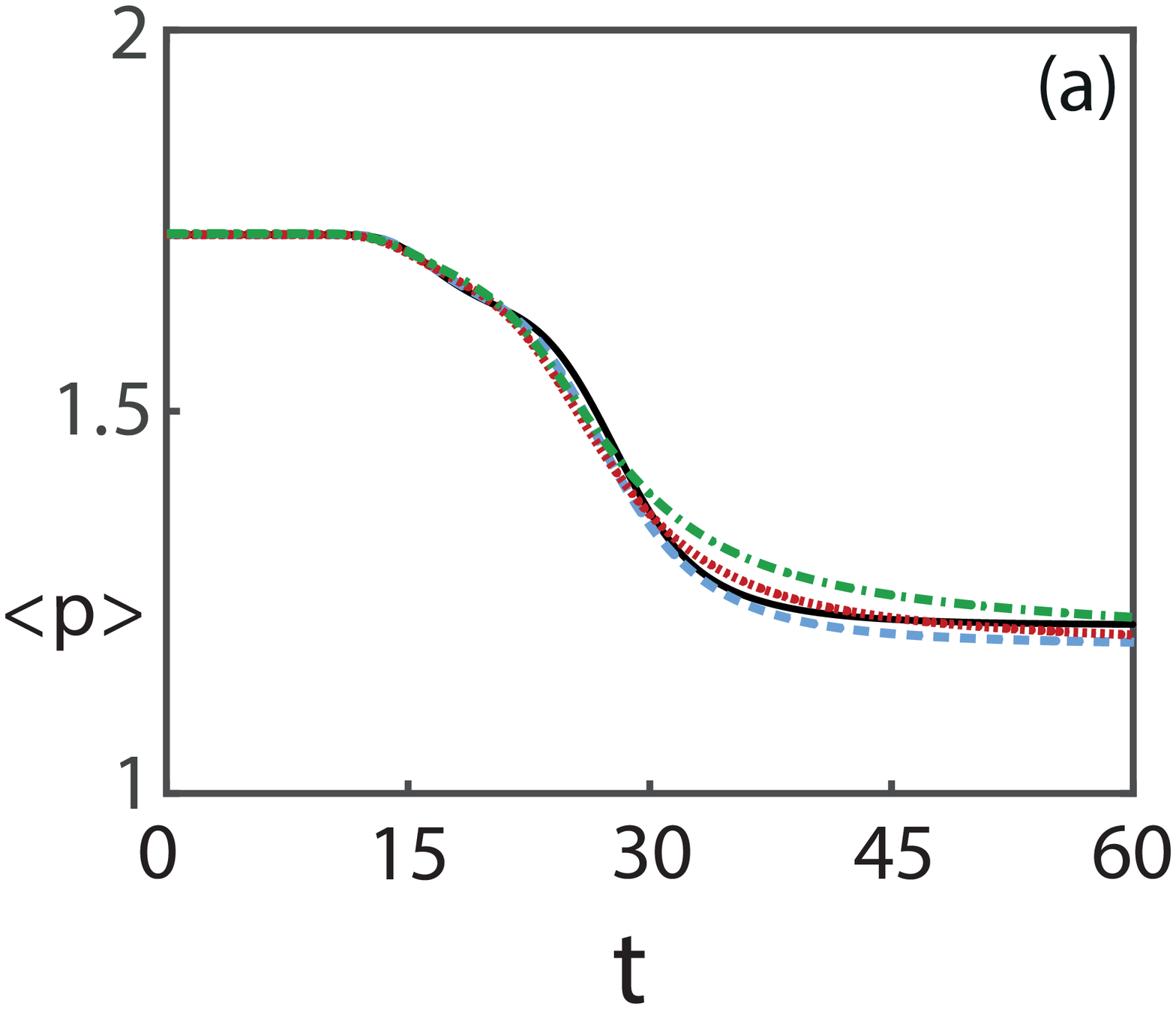,scale=0.24}\hspace*{0.1 cm}\epsfig{file=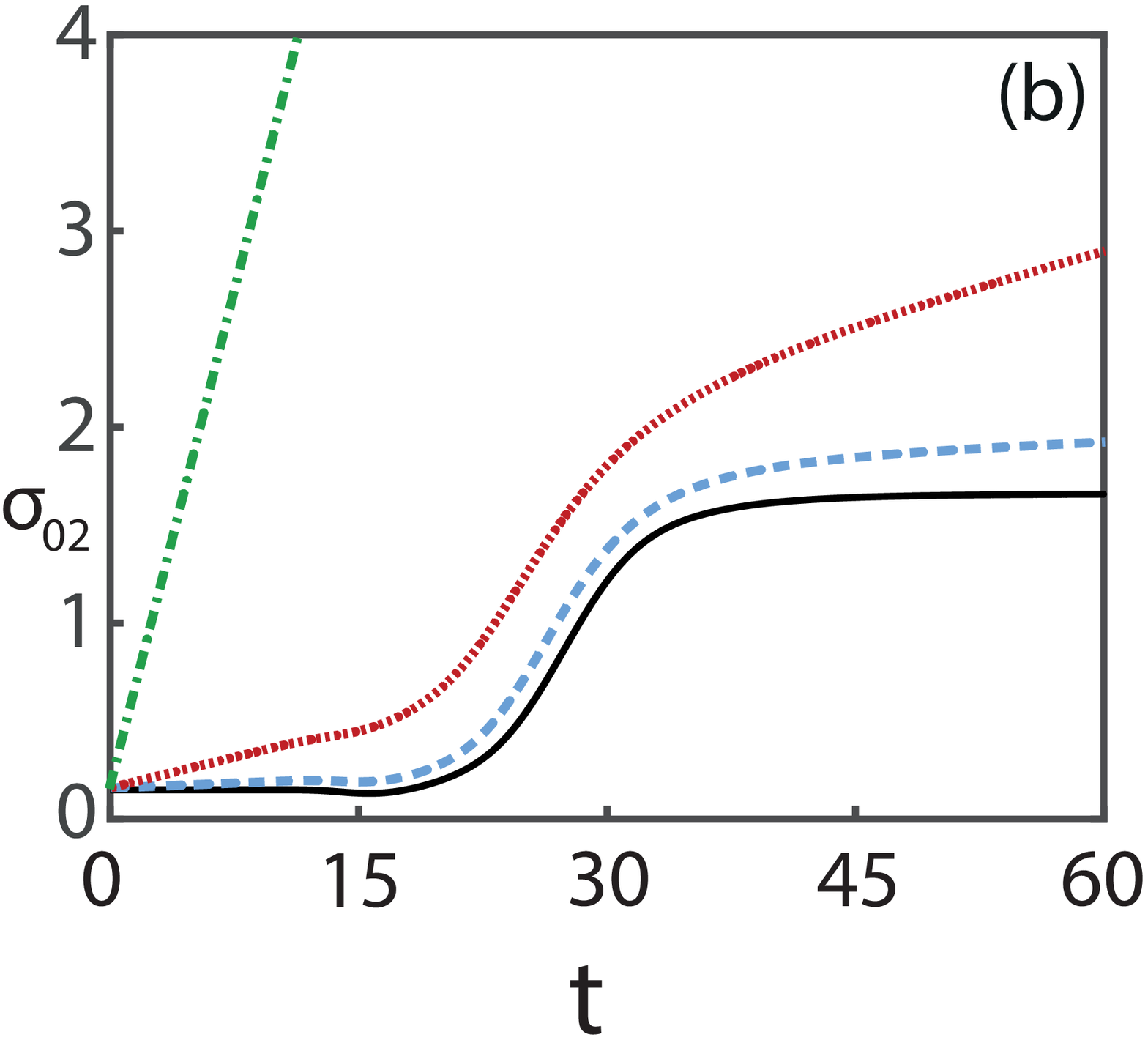,scale=0.24} 
\caption{Average momentum $<p>$ (a) and momentum spread $\sigma_{02}$ (b) in dimensionless units as functions of time for $E_K=1.5$, $\tau=3$, and $\sigma_0=0.1E_K$. The values of $\lambda$ are the same as in Fig. \ref{fig:dens05}. }\label{fig:pav15}
\end{center}
\end{figure}

\begin{figure}[ht!]
\setlength{\unitlength}{1cm}
\begin{center}
\epsfig{file=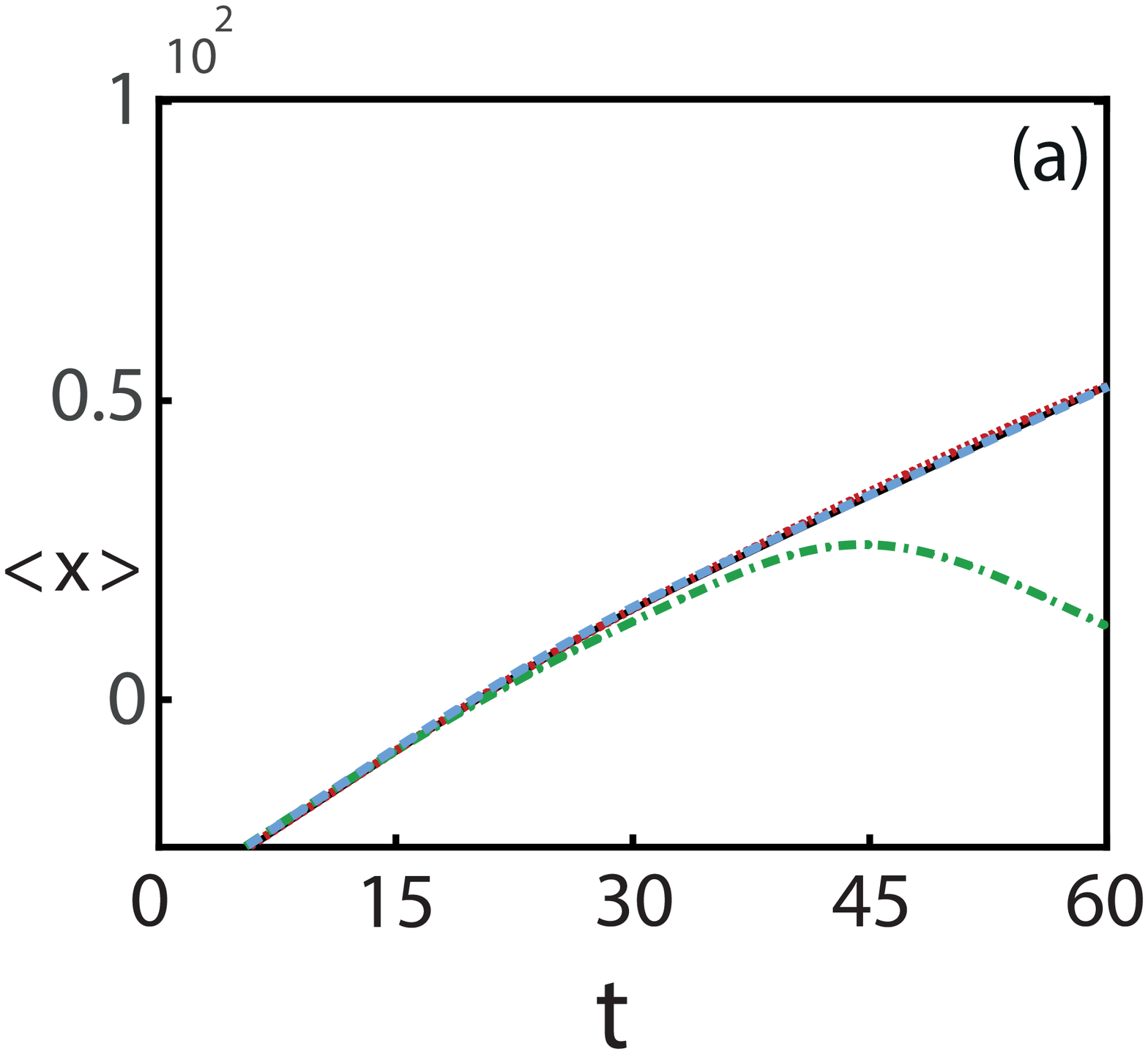,scale=0.24}\hspace*{0.1 cm}\epsfig{file=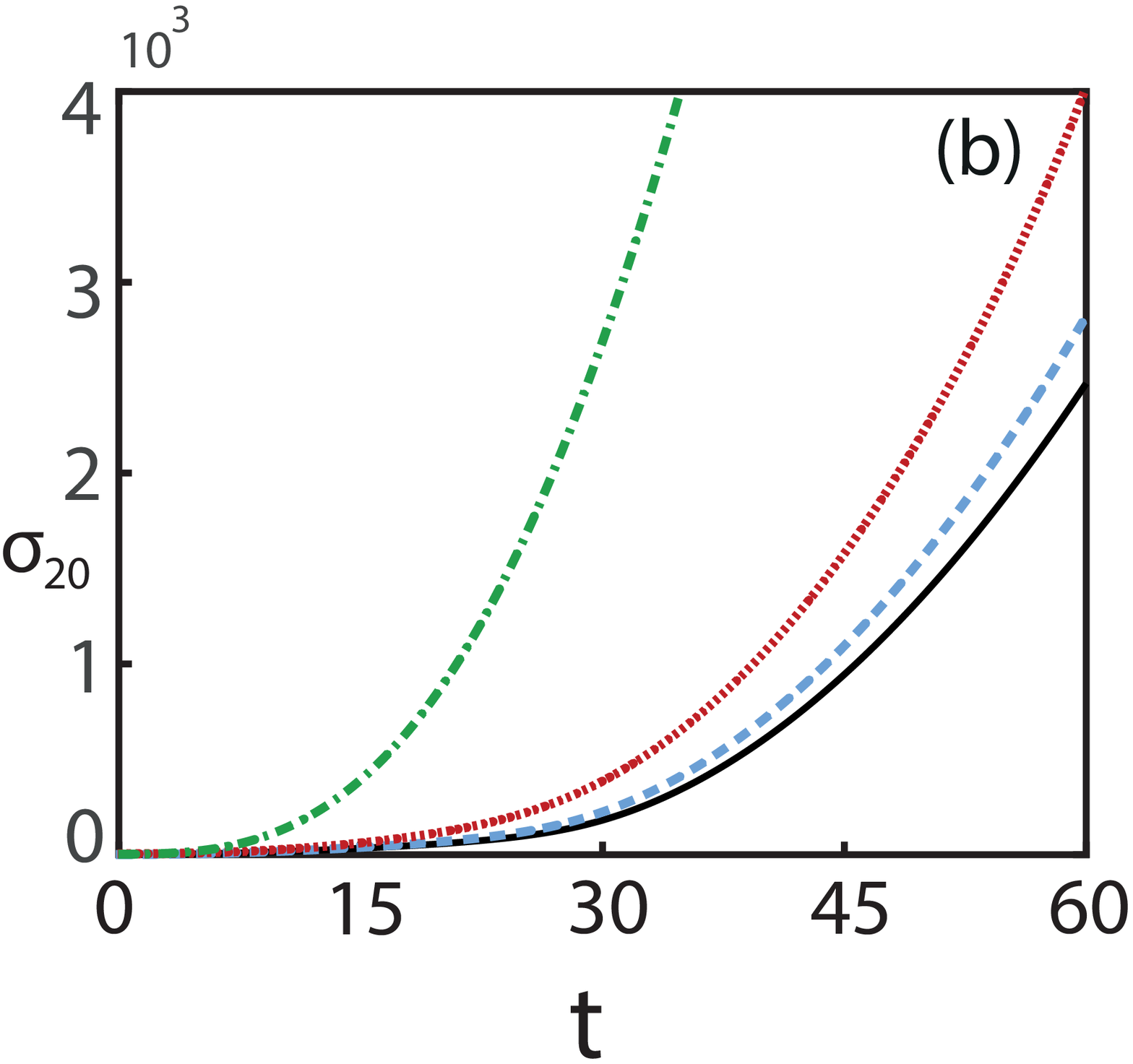,scale=0.24} 
\caption{Average position $<x>$ (a) and position spread $\sigma_{20}$ (b) in dimensionless units as functions of time for $E_K=1.5$, $\tau=3$, and $\sigma_0=0.1E_K$. The values of $\lambda$ are the same as in Fig. \ref{fig:dens05}. }\label{fig:xav15}
\end{center}
\end{figure}

\begin{figure*}[ht!]
\begin{center}
\epsfig{file=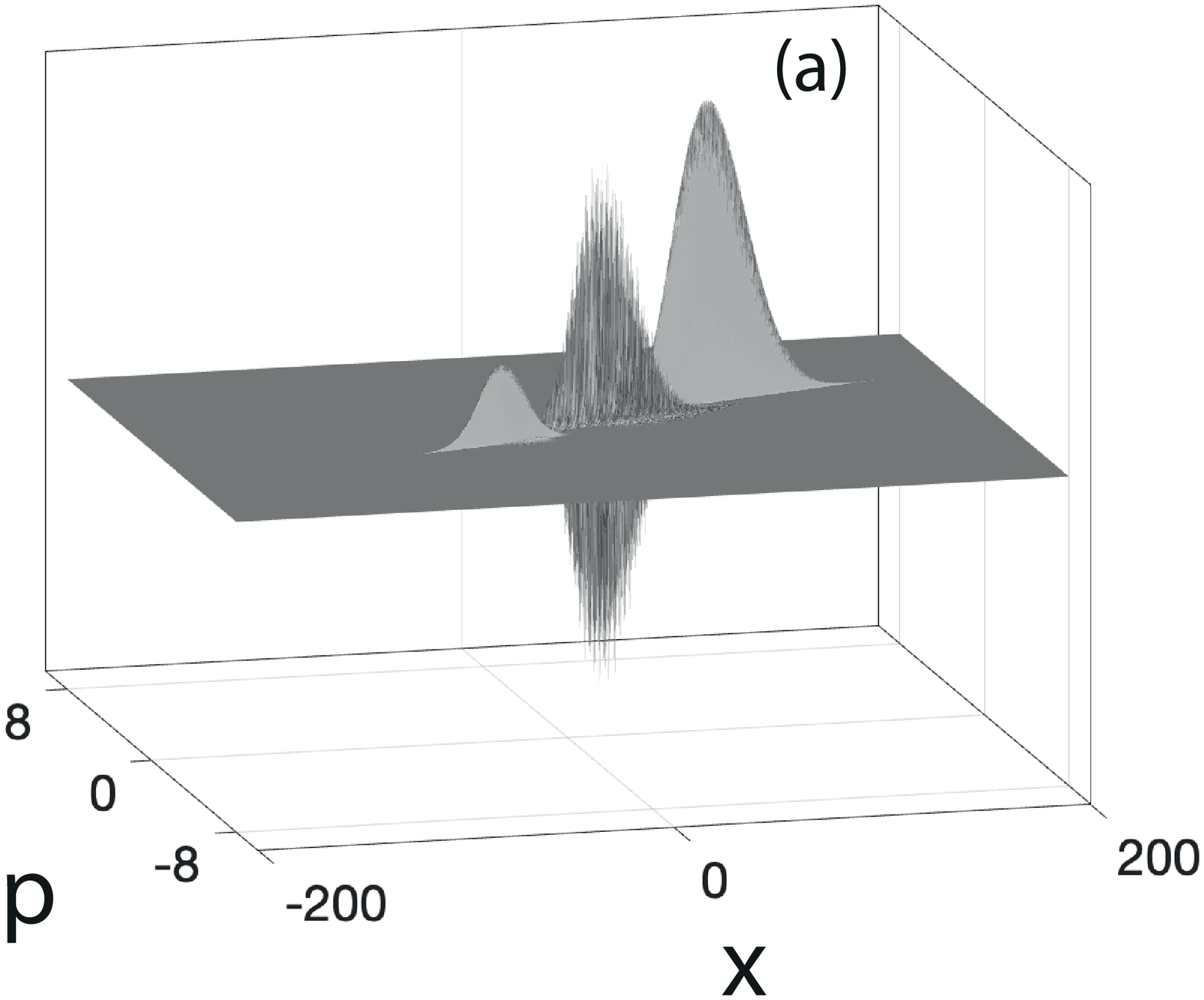,scale=0.25}\epsfig{file=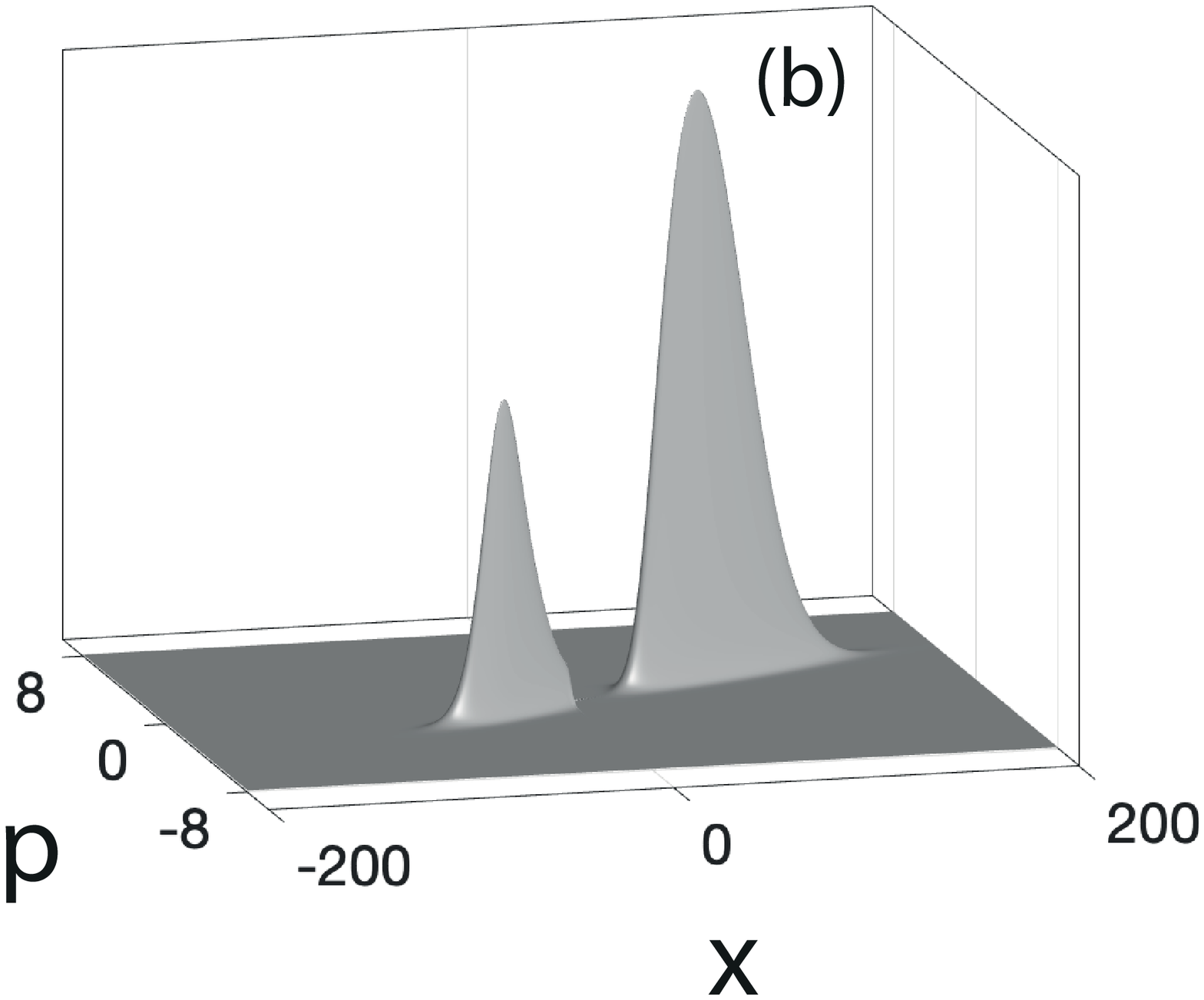,scale=0.3} 
\epsfig{file=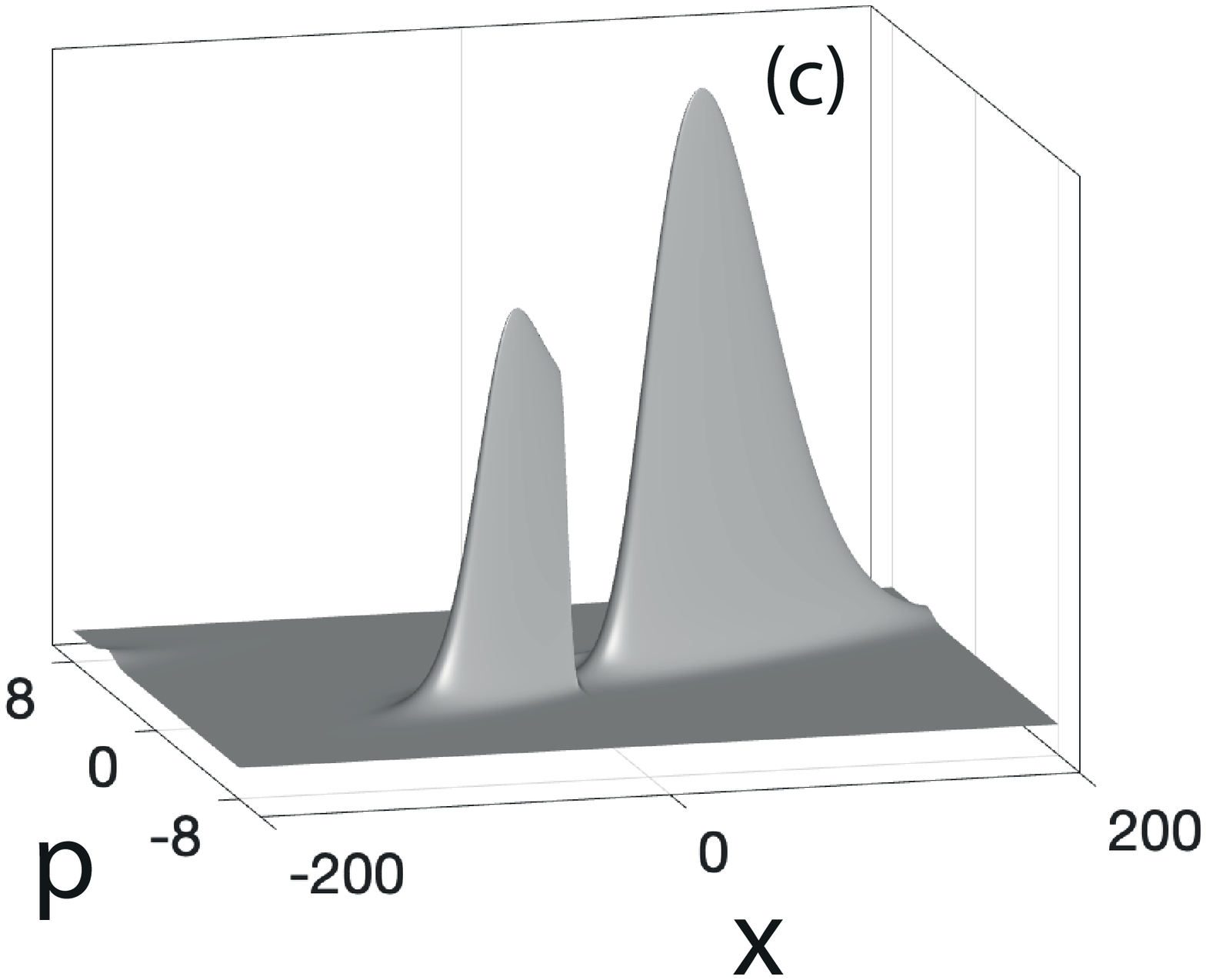,scale=0.3}\caption{Wigner function $f(x,p)$ at $t=60$, for $E_K=1.5$, $\tau=3$, and $\sigma_0=0.1E_K$. The values of $\lambda$ are the same as in Fig. \ref{fig:wig05}.} \label{fig:wig15}
\end{center}
\end{figure*}

Finally, we illustrate the scattering dynamics for an initial energy above the potential barrier, $E_K=1.5$. The density profiles are shown in Figs. \ref{fig:dens15}(a)-\ref{fig:dens15}(d) at the same instants of time and for the same values of $\lambda$ as in the previous cases. The average momentum and the momentum spread are shown in Figs. \ref{fig:pav15}(a) and \ref{fig:pav15}(b), the average position and the position spread in Figs. \ref{fig:xav15}(a) and \ref{fig:xav15}(b), and the Wigner function in Figs. \ref{fig:wig15}(a)-\ref{fig:wig15}(c). 


The initial energy is now higher than the potential one. 
By comparing Figs. \ref{fig:pav15} and \ref{fig:xav15} for the average momentum and the average position with the corresponding Figs. \ref{fig:pav05} and \ref{fig:xav05}, we see that the effect of a decreasing $\lambda$ on transmission is now much weaker;
the average momentum, after the drop from the initial value during the interaction, settles to a value which depends very little on the coherence length $\lambda$. 
The same is true for the average position [see Fig. \ref{fig:xav15}(a)], but only for the long and intermediate values of $\lambda$, the unphysical behavior for $\lambda=1$ being present in this case as well [see the dash-dotted green line in Figs. \ref{fig:pav15} and \ref{fig:xav15}]. The density profiles show a similar behavior as in the previous two cases, with the onset of oscillations during the interaction of the packet with the potential, followed by a separation into reflected and transmitted portions, which become lower and broader as the value of  $\lambda$  is reduced. 
\subsection{The transmission coefficient.}
\begin{figure}[ht!]
\setlength{\unitlength}{1cm}
\begin{center}
\epsfig{file=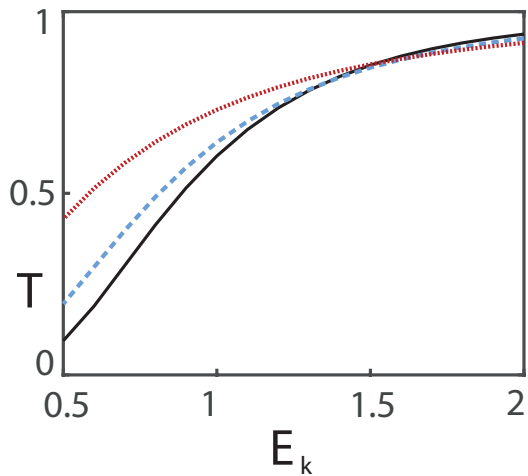,scale=0.4}
\caption{Transmission coefficient $T$ as a function of the energy for $0.5 \le E_K \le 2$, $\tau=3$, and three different values of $\lambda$. Coherent case ($\lambda\rightarrow \infty$): solid black line; $\lambda=10$: dashed blue line; $\lambda=4$: dotted red line.}\label{fig:Transmcoeff}
\end{center}
\end{figure}

Finally, in Fig. \ref{fig:Transmcoeff} we report the transmission coefficient, as given by Eq. (\ref{eqn:T}) as a function of the energy for $\tau=3$ and three values of the correlation length, $\lambda\rightarrow \infty$, $\lambda=10$, and $\lambda=4$. We see that, at energies below the potential barrier, the transmission coefficient increases as the correlation length $\lambda$ is made smaller (the dashed blue and dotted red curves, corresponding to $\lambda=10$ and $\lambda=4$, respectively, lie above the solid black curve, which represents the results of the coherent case), while this tendency is reversed  at energies above the potential barrier. This shows that decoherence (low values of $\lambda$) favors transmission ($T$ values are incremented) in reflection-dominated regimes, while it favors reflection at high energy. The decoherence effects at high energies also appear weaker than at low energies (the curves of the transmission coefficient lie closer to each other at the higher values of $E_K$).  This observation is consistent with the spreading effect of the finite correlation length on the Wigner function, which enhances the size of the transmitted packet at lower energies and the size of the reflected packet at higher energies. We have also calculated the curves of the transmission coefficient for other values of $\tau$ in the vicinity of $\tau=3$ with no relevant deviations from the ones shown in Fig. \ref{fig:Transmcoeff}. \\

\subsection{The density jump.} \label{sub:Discontinuity}

In all three cases examined in our simulations, the Wigner function and, as a consequence, the density profile exhibit a sudden jump, which appears as a discontinuity at $x=0$ in our graphs. This feature has never been observed in the Wigner function simulations performed by us or by other groups. We have analyzed this phenomenon by varying the width of the potential $a$ [see Eq. (\ref{potential})] and by adopting a top view of the Wigner function. In all simulations performed so far the choice $a=1$ was done, in accordance with the adimensionality of the $x$ variable. \\

\begin{figure}[ht!]
\begin{center}
\epsfig{file=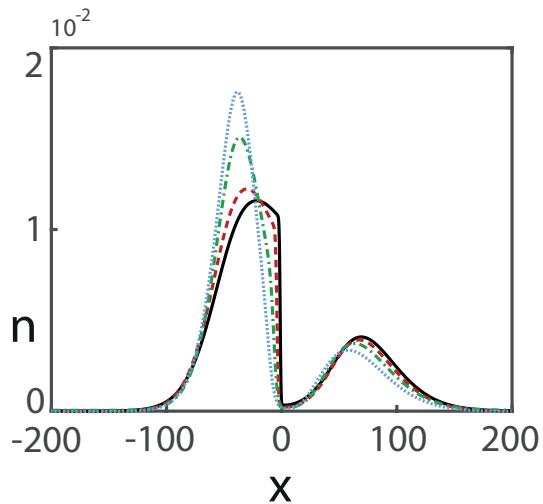,scale=0.4} 
\caption{Density $n(x)$ in dimensionless units as a function of $x$ at $t=60$ for $E_K=0.5$, $\tau=3$, $\sigma_0=0.1E_K$, and $\lambda = 4$ for four values of the potential width: solid black line: $a=1$, dashed red line: $a=2$, dash-dotted green line: $a=5$, and dotted blue line: $a=8$.} 
\label{fig:variaa}
\end{center}
\end{figure}

In Fig. \ref{fig:variaa} we show the density profile $n(x)$ at $t=60$, for $E_K=0.5$, $\tau=3$, $\sigma_0=0.1E_K$, and $\lambda = 4$ for several values of the potential width $a$: the solid black line refers to $a=1$, the dashed red line to $a=2$, the dash-dotted green line to $a=5$, and the dotted blue line to $a=8$. We see that, as the potential is made broader, the density jump is smoothed out and eventually the profile becomes regular. This indicates that what appears as a discontinuity at $a=1$ is in fact a narrow region of rapid decrease in the profile, not well resolved by the numerical discretization of the phase-space variables. 

\begin{figure}[ht!]
\begin{center}
\epsfig{file=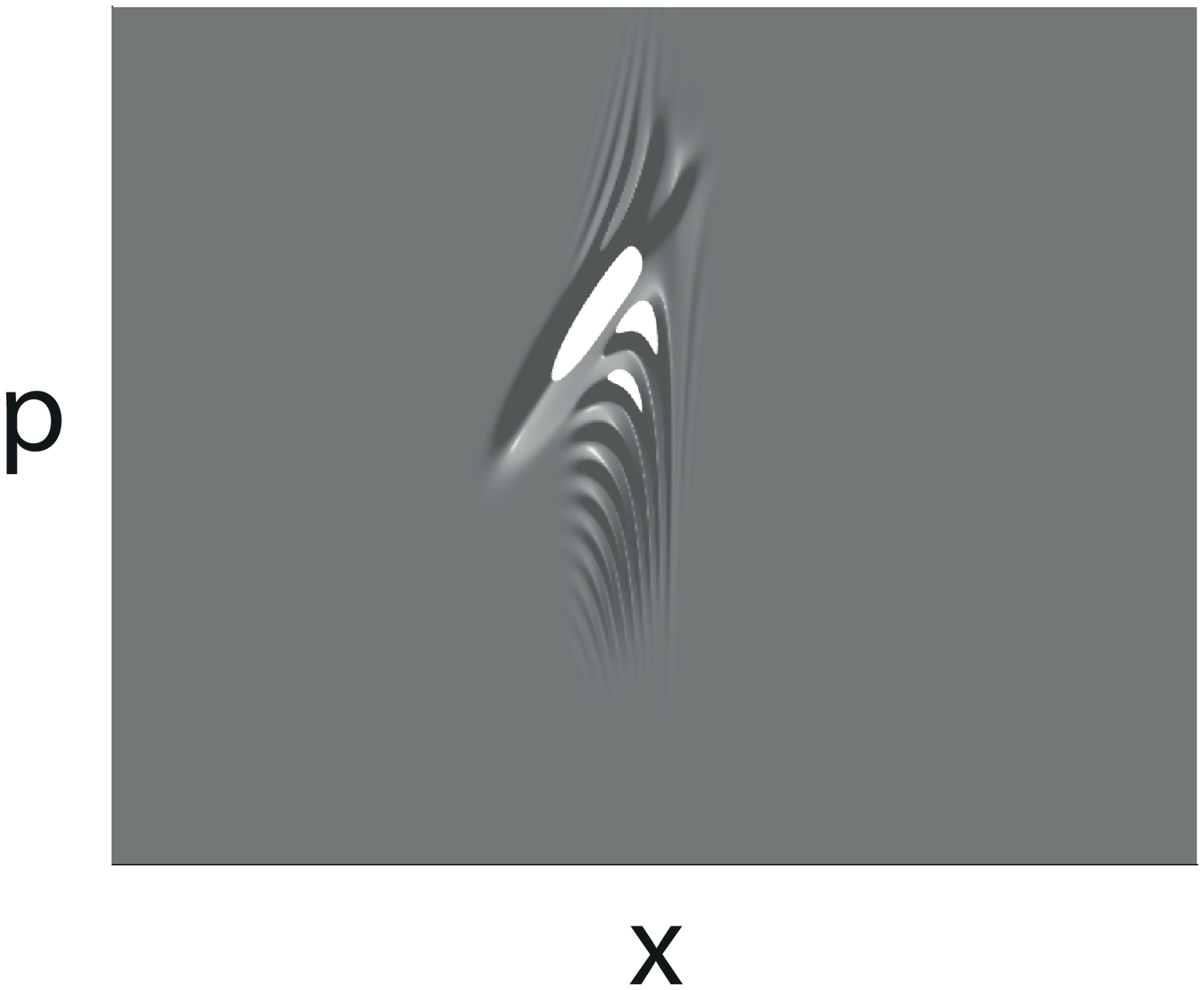,scale=0.21} 
\epsfig{file=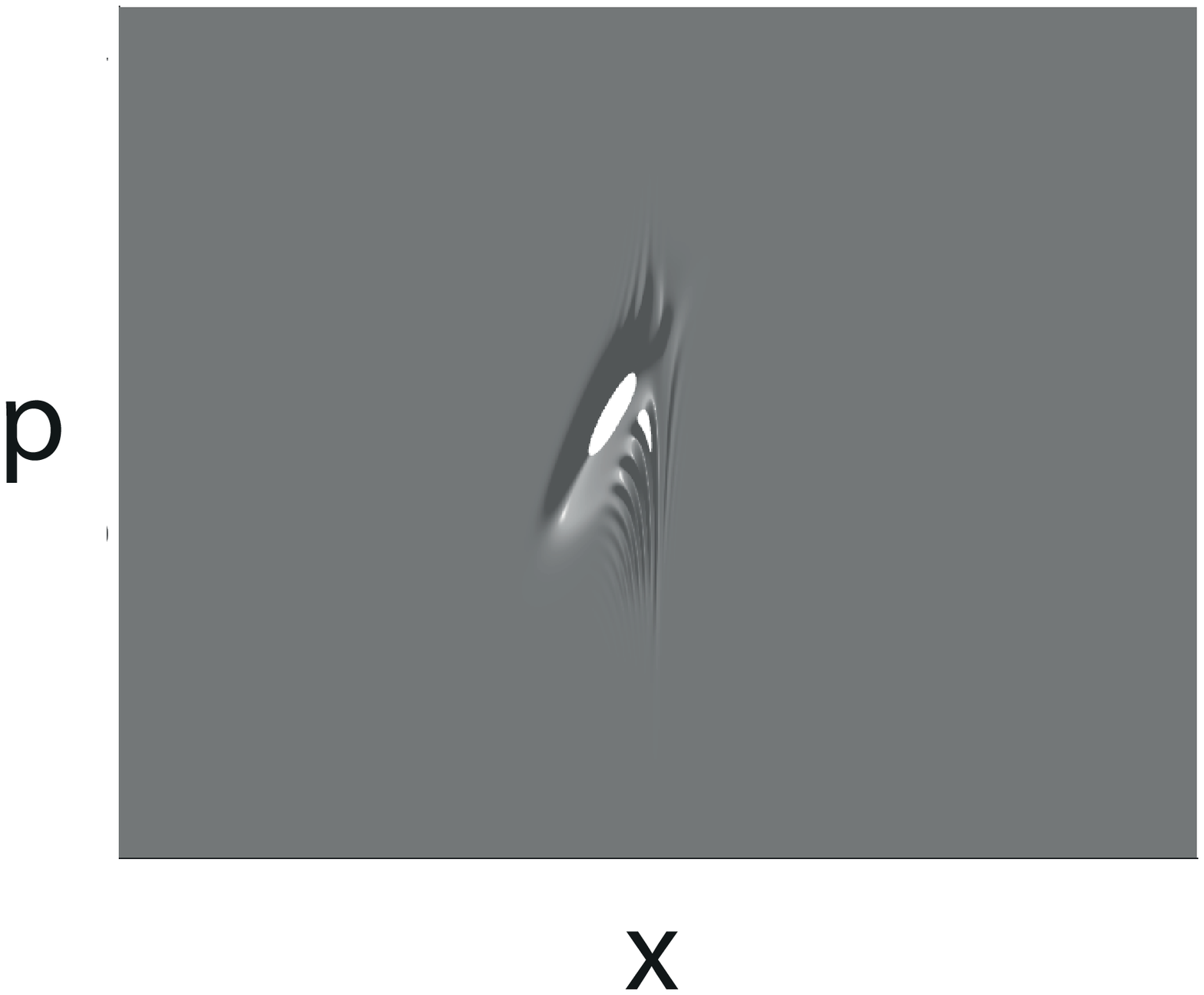,scale=0.21} 
\epsfig{file=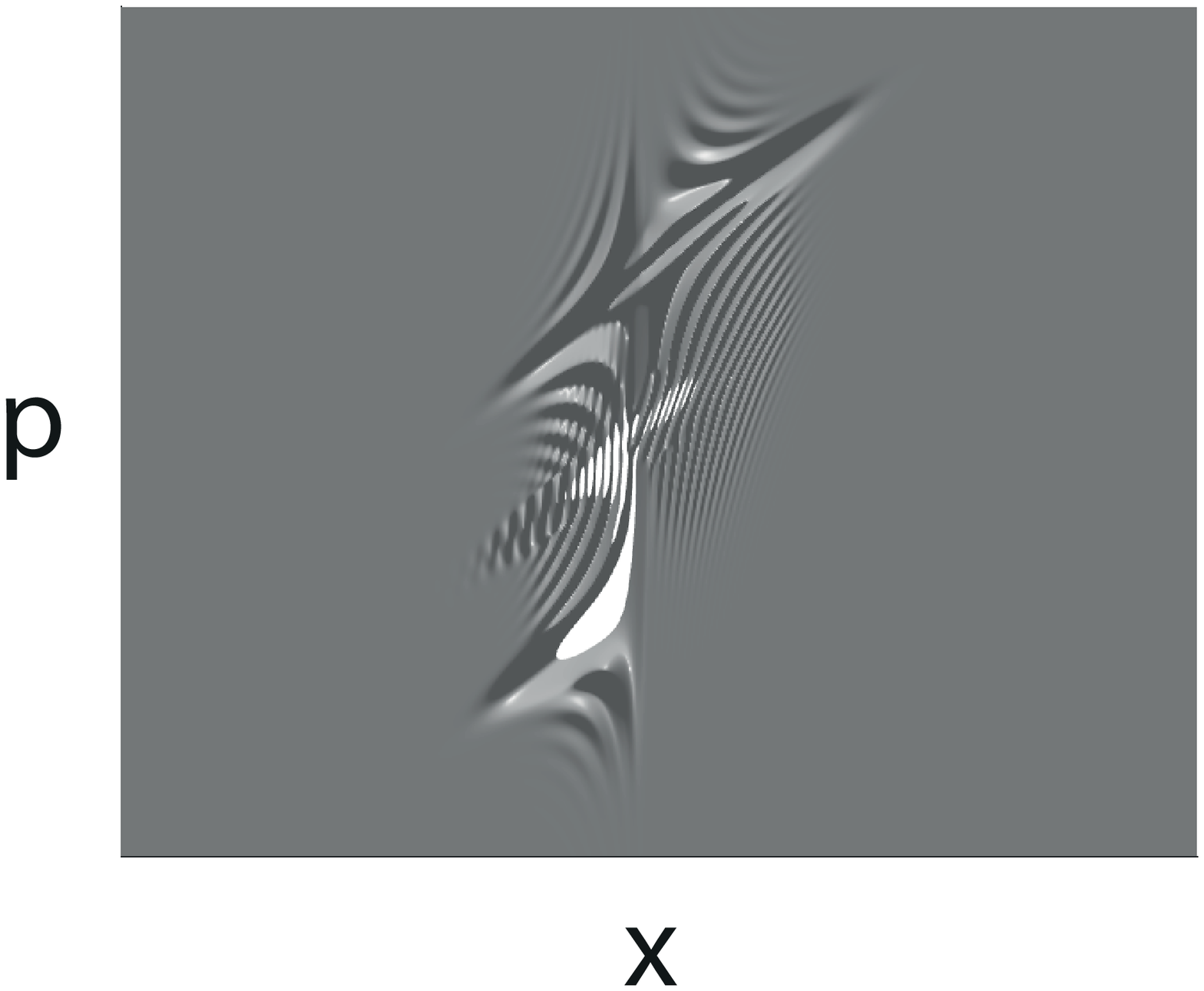,scale=0.21} 
\epsfig{file=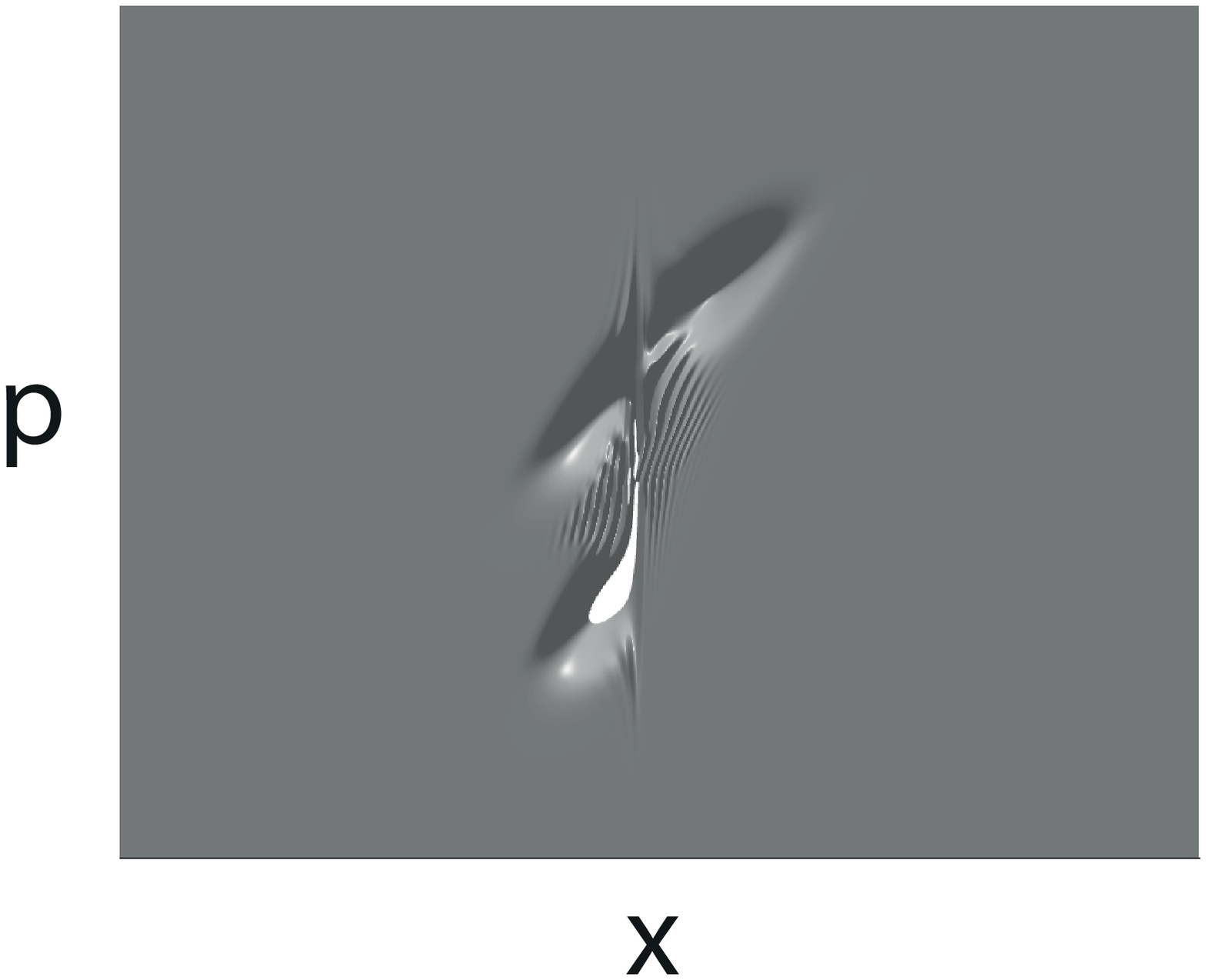,scale=0.21}
\epsfig{file=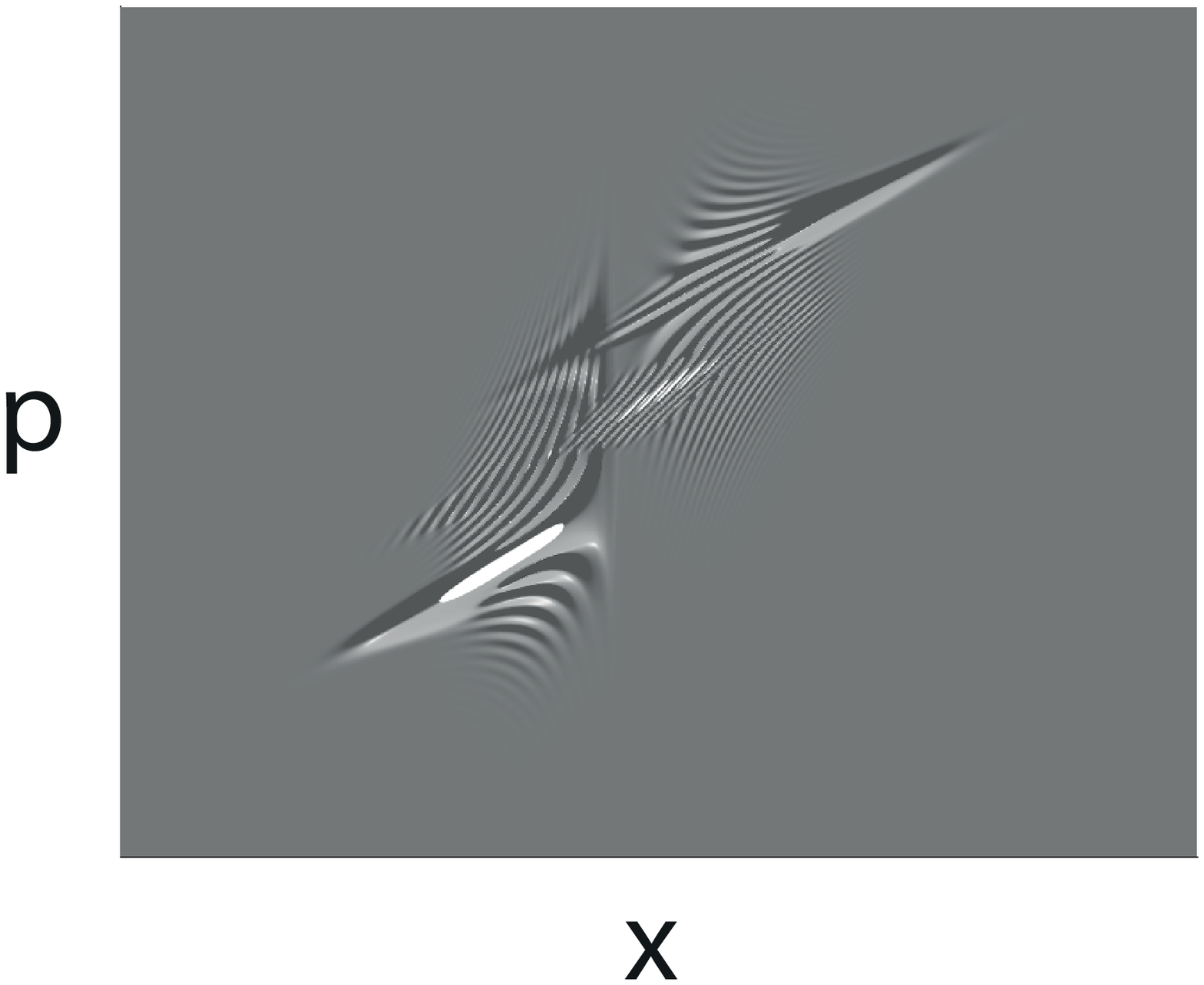,scale=0.21}  
\epsfig{file=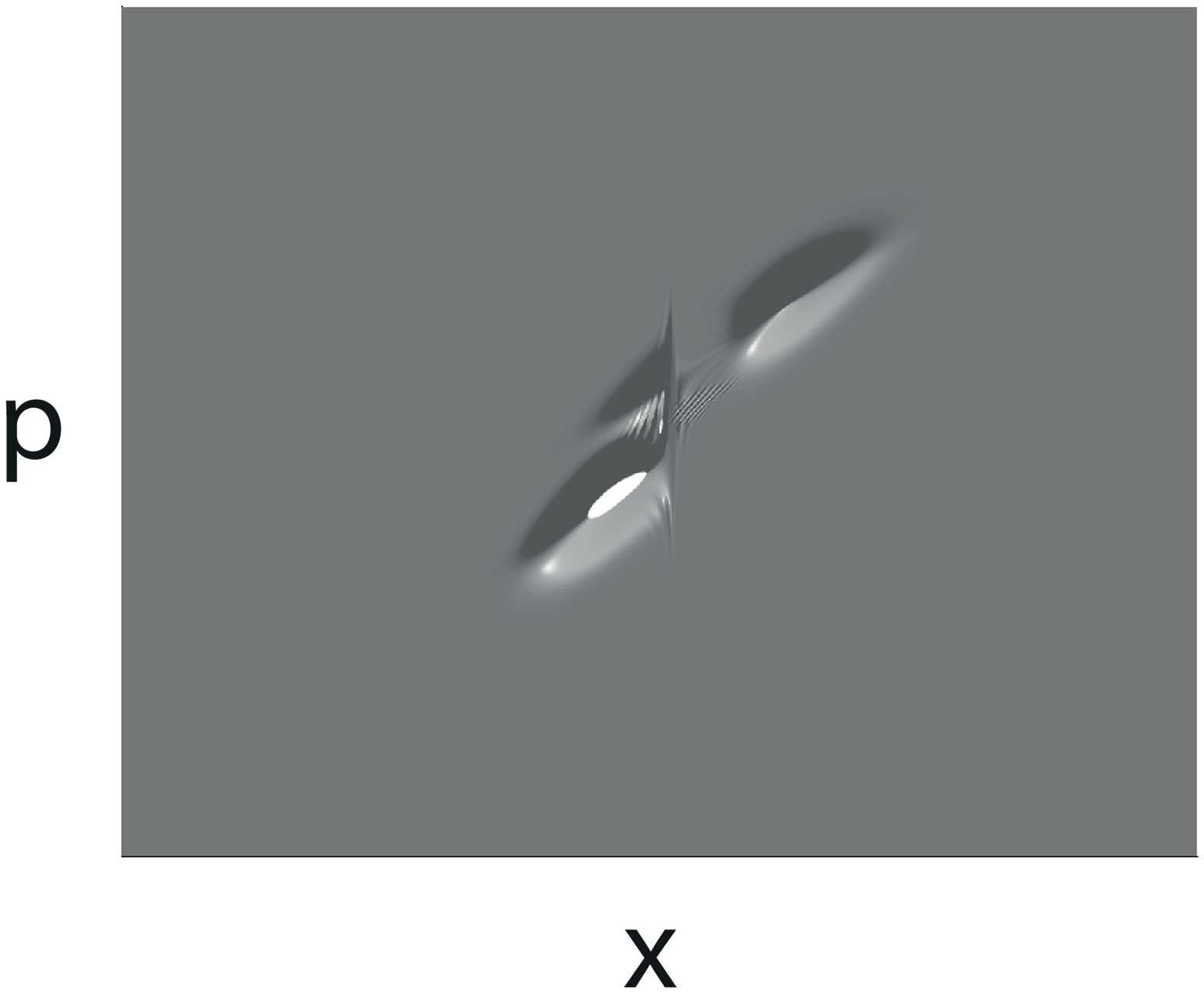,scale=0.21}
\epsfig{file=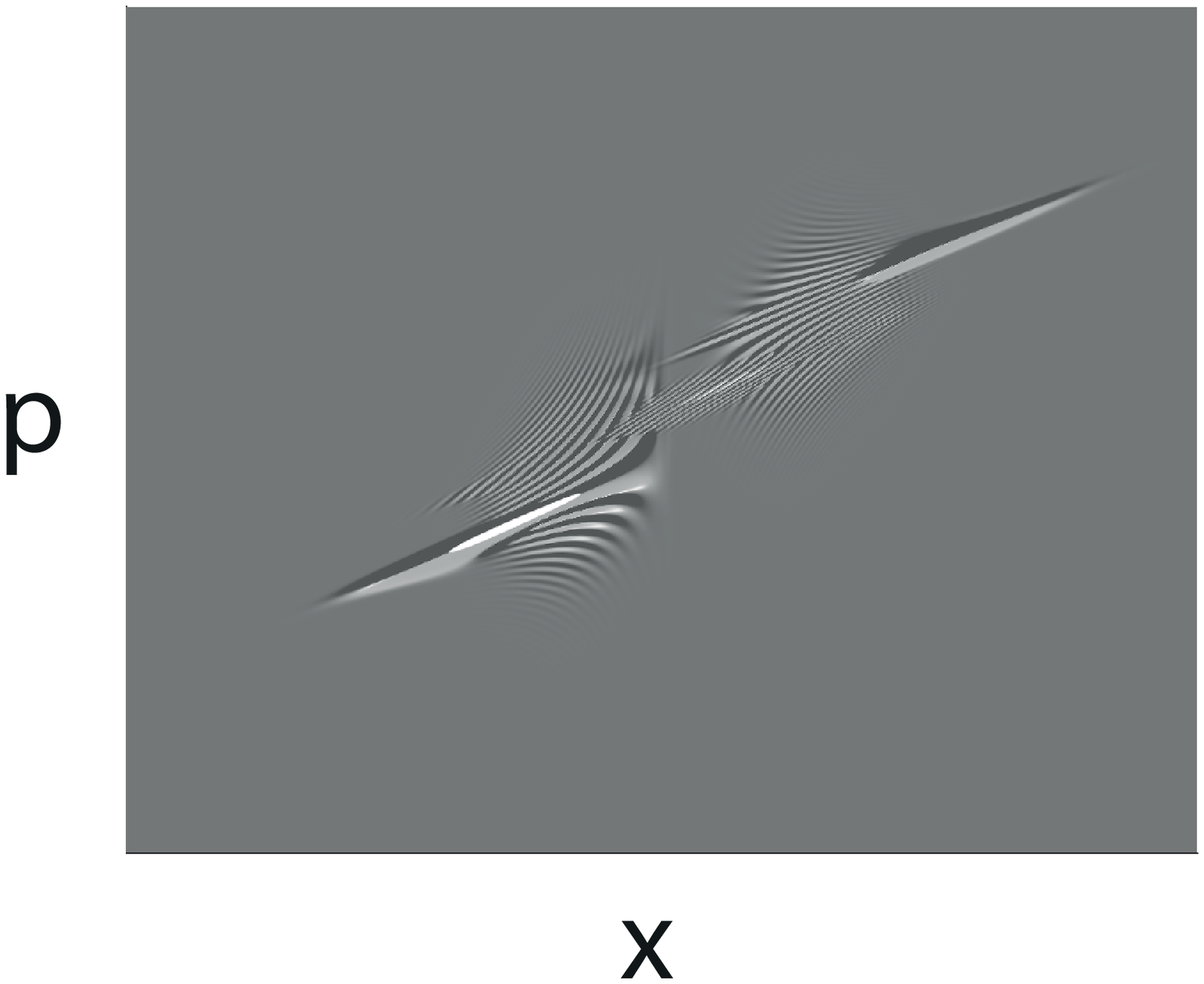,scale=0.21}
\epsfig{file=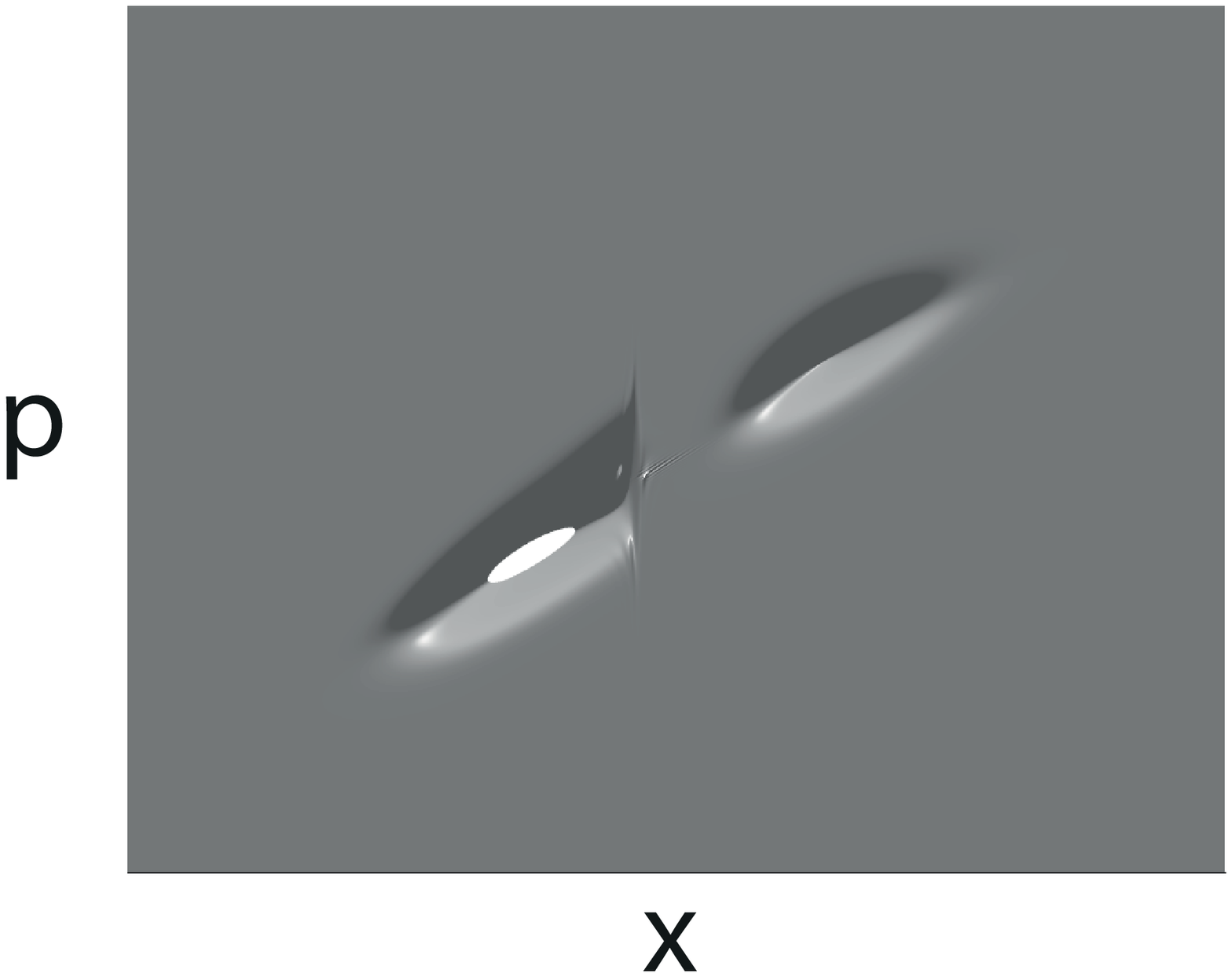,scale=0.21}
\epsfig{file=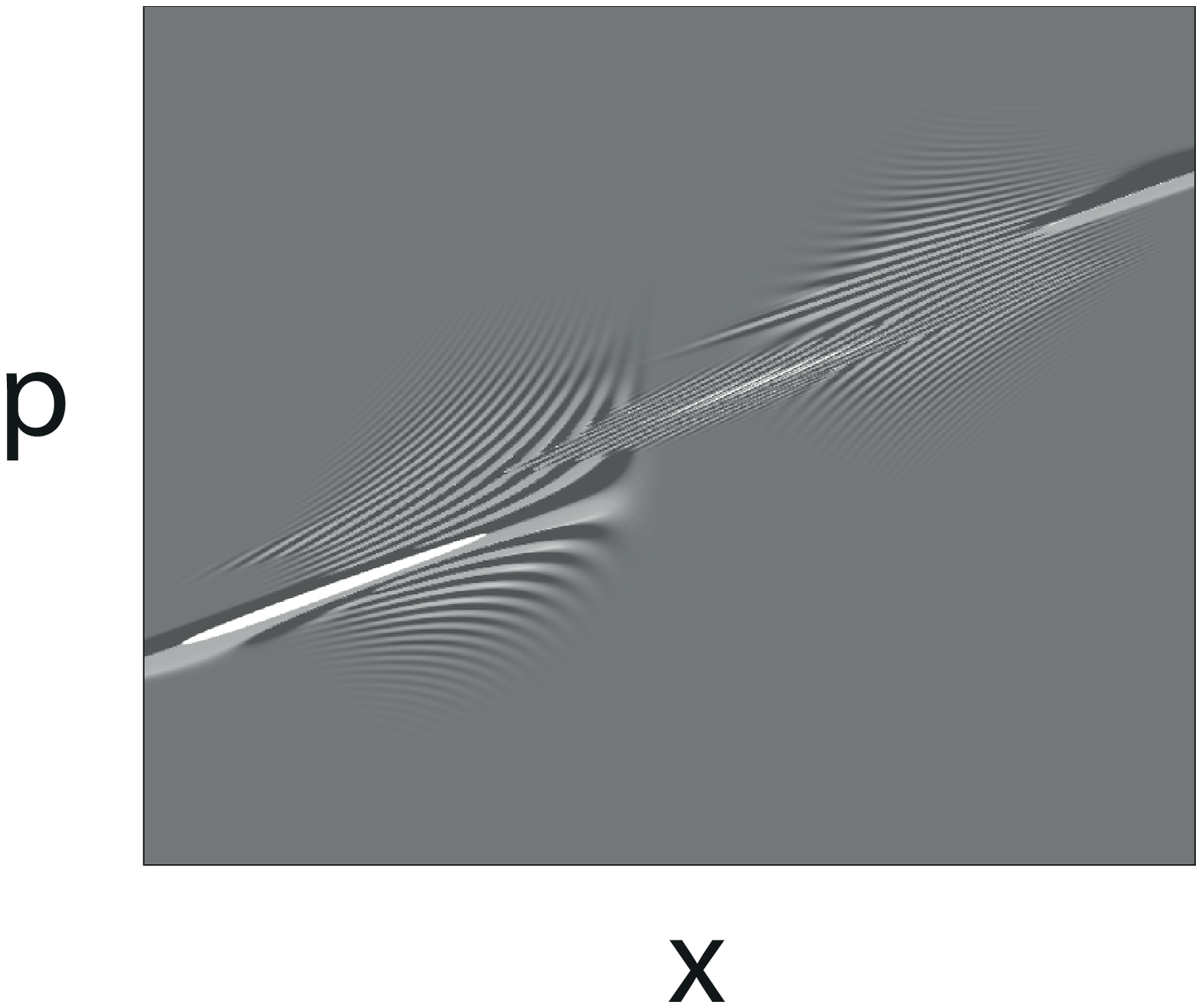,scale=0.23} %\hspace*{0.1cm}
\epsfig{file=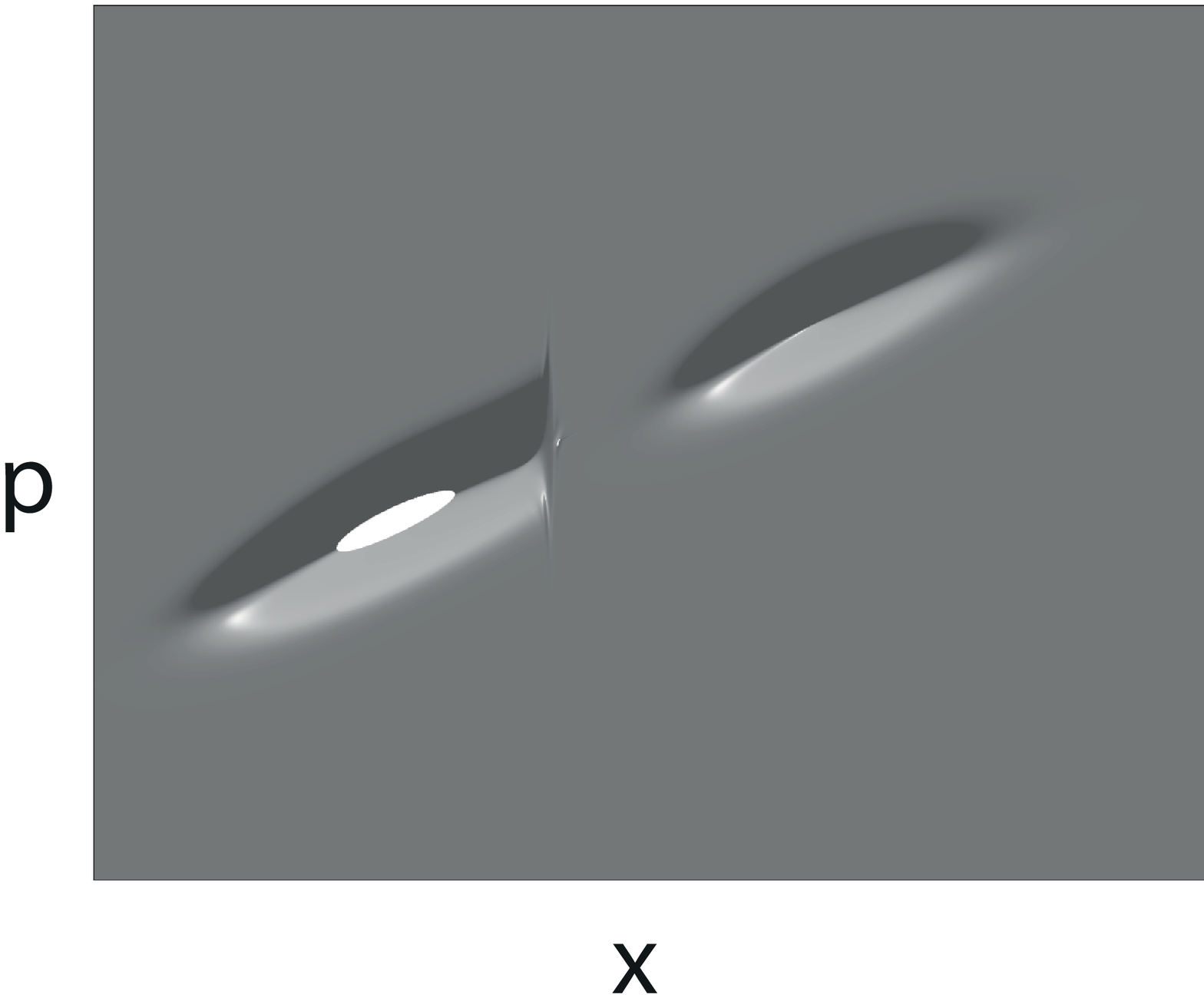,scale=0.18} 
\caption{Wigner function $f(x,p)$ as a function of $x$ and $p$ at five instants of time,  $t=12$, $t=24$, $t=36$, $t=48$ and $t=60$ for $E_K=0.5$, $\tau=3$, and $\sigma_0=0.1E_K$; left panels: coherent case ($\lambda\rightarrow \infty$); right panels: $\lambda = 10$.} 
\label{fig:wignertop}
\end{center}
\end{figure}

We explain this phenomenon with the aid of Fig. \ref{fig:wignertop}, which shows a top view of the Wigner function at five instants of time, $t=12$, $t=24$, $t=36$, $t=48$, and $t=60$ for $E_K=0.5$, $\tau=3$, $\sigma_0=0.1E_K$, and $\lambda\rightarrow \infty$ (left panels: coherent case) and $\lambda = 10$ (right panels). Let us consider for a moment the classical picture: if the initial distribution were to evolve classically, it would follow the phase-space trajectories, the bulk being reflected back and turning from the $p>0$ to the $p<0$ half-plane, with only a small portion in the tail being transmitted. The non locality due to the quantum effects in the coherent case makes the distribution deviate from the classical trajectories and causes the interference fringes seen as oscillations on the Wigner function in the potential region (see Fig. \ref{fig:wignertop}, left panels). The finiteness of the correlation length switches off the exchange of information and the non-locality effects on a spatial scale of order $\lambda$, leaving a void central region of size $\sim \lambda$ (see Fig. \ref{fig:wignertop}, right panels).  The fact that the finite correlation length somehow restores  locality into the system does not mean that the system is becoming entirely classical, since, for example, the quantum character of the pseudo-differential operator describing the interaction with the potential remains in place.

\section{Conclusions.} \label{sec:Conclusions}
In this paper we have investigated numerically the decoherence model introduced in the WF formalism in Ref. \cite{Barletti18}. We have considered a simple system given by an initial Gaussian Wigner function undergoing scattering by a Gaussian potential barrier. Three energy regimes were analyzed, a reflection-dominated regime with energy below the barrier height, a transmission--dominated regime with energy above the barrier height and an intermediate regime with energy equal to the barrier height. The effects of the decoherence mechanism were studied by considering three different values of the correlation length $\lambda$, and compared with the coherent case. The main conclusion is that of an enhanced transmission at low energies and a reduced transmission at higher energies, the former effect being more pronounced, also because a finite correlation length favors
transmission of low energy electrons through the potential
barrier, inhibiting reflection,
since long wavelength components of the potential
cannot interfere effectively with the electron wave
function. We interpret this behavior as due to the broadening and flattening of the Wigner function as $\lambda$ is reduced, which causes an increased size of the transmitted packet at small energies and of the reflected packet at high energies. Finally, as the coherence length is reduced, the Wigner function exhibits a narrow region of sharp variation, which appears when the packet separation begins; our analysis strongly suggests that this is due to the tendency of the system to show a more classical behavior at low values of the correlation length.

\end{document}